\documentstyle[emulateapj,fleqn]{article}
\begin{document}
\title{Jet Acceleration by Tangled Magnetic Fields} \author{Sebastian
  Heinz\altaffilmark{1} \and Mitchell C. Begelman\altaffilmark{2,3}}
\affil{JILA, University of Colorado and National Institute for Standards
  and Technology, Boulder, Colorado 80309-0440\altaffilmark{4}}
\altaffiltext{1}{email address: heinzs@rocinante.Colorado.edu}
\altaffiltext{2}{email address: mitch@jila.Colorado.edu}
\altaffiltext{3}{also at Institute for Theoretical Physics, University of
  California, Santa Barbara}
\altaffiltext{4}{also at Department of Astrophysical and Planetary
  Sciences, University of Colorado, Boulder}


\begin{abstract}
  We explore the possibility that extragalactic radio jets might be
  accelerated by highly disorganized magnetic fields that are strong enough
  to dominate the dynamics until the terminal Lorentz factor is reached.
  Following the twin-exhaust model by Blandford \& Rees (1974), the
  collimation under this scenario is provided by the stratified thermal
  pressure from an external medium.  The acceleration efficiency then
  depends on the pressure gradient of that medium.  In order for this
  mechanism to work there must be continuous tangling of the magnetic
  field, changing the magnetic equation of state away from pure flux
  freezing (otherwise conversion of Poynting flux to kinetic energy flux is
  suppressed).  This is a complementary approach to models in which the
  plasma is accelerated by large scale ordered fields.  We include a simple
  prescription for magnetic dissipation, which leads to tradeoffs among
  conversion of magnetic energy into bulk kinetic energy, random particle
  energy, and radiation.  We present analytic dynamical solutions of such
  jets, assess the effects of radiation drag, and comment on observational
  issues, such as the predicted polarization and synchrotron brightness.
  Finally, we try to make the connection to observed radio galaxies and
  $\gamma$-ray bursts.
\end{abstract}


\section{Introduction}
Many extragalactic radio jets move with bulk Lorentz factors $\Gamma > 10$,
as evidenced by very short variability timescales, superluminal proper
motion of jet features, and dilated particle lifetimes, yet the process
that actually accelerates the material is not well known, and neither is
the mechanism that collimates the outflow.

One of the first serious models for the large scale dynamics of
extragalactic radio jets is the `twin exhaust' model (Blandford \& Rees 74,
BR74 throughout the rest of the paper).  In this model, relativistic
particle pressure provides the bulk acceleration via conversion of internal
to kinetic energy.  The collimation comes from confinement by an external
medium, the pressure of which is very likely stratified in the
gravitational field of the central black hole and the host galaxy and
cluster (either in a hydrostatic equilibrium configuration or as a wind).
Pressure balance then dictates the evolution of the jet bulk Lorentz factor
$\Gamma$, and the jet diameter $R$.  However, cooling processes for
particles with highly relativistic random motions (necessary to produce
outflows with large bulk Lorentz factors $\Gamma$) are very efficient, thus
competing with and probably disabling bulk acceleration.

The problem of radiative losses is similarly limiting in the case of
radiatively accelerated jets (Phinney 1982).  Furthermore, these jets rely
on the presence of a strong point-like radiation source (the terminal
Lorentz factor is limited by the solid angle subtended by the radiation
source due to relativistic aberration).  The same Inverse Compton (IC)
scattering effects that lead to the bulk acceleration lead to even stronger
radiative losses.  In fact, radiation drag can actually decelerate the jet
if the external radiation field is isotropic, which leads to both random
energy losses {\it and} kinetic energy losses by the same process.

Radiative losses are much less limiting if the energy is stored in the
magnetic field.  Organized magnetic fields have been suggested to provide
both acceleration and collimation (e.g., Blandford \& Payne 1982, Li,
Chiueh, \& Begelman 1992), however, such solutions, in which the
collimation is provided by toroidal field, seem to be hampered by
instabilities (Begelman 1998).  Furthermore, it is difficult to achieve
collimation on the basis of magnetic tension alone without contribution
from an external pressure source (Begelman 1995, B95 hereafter).

Polarization measurements show that the magnetic field in jets is probably
{\it not} well organized --- the polarization is generally well below the
maximal value of $\sim 70$\%; only in knots does the polarization tend
towards this value (note, however, that interpretations of polarization
measurements are often ambiguous, since different field geometries can
sometimes lead to the same net polarization).  This argues for the presence
of largely unorganized, chaotic fields, which could easily account for the
high polarization measured in the knots if they are interpreted as shocks,
compressing the field in the shock plane (Laing 1980, B95).  This goes hand
in hand with the fact that the field produced in the disk by dynamo
processes is expected to be highly chaotic.  Since the conditions in the
jet will likely be controlled by disk physics, we should expect the same
statement to be true for the magnetic field in the jet at least close to
the disk.  These arguments led us to investigate the dynamics of jets
containing large amounts of such disorganized magnetic fields.

The rate of acceleration in a jet propelled by internal (isotropic)
particle pressure in an external pressure gradient is limited to $\Gamma
\propto {p_{\rm ext}}^{-1/4}$ (BR74), which means that the acceleration to
bulk Lorentz factors of $\Gamma \sim 10 - 100$ would occur over length
scales $\gtrsim (1000)\, r_{\rm g}$ for external pressure gradients $p
\propto z^{-2}$.  One might think that an anisotropic pressure in the form
of chaotic magnetic fields could increase the rate at which the jet is
accelerated, if the excess momentum flux is oriented along the direction of
the jet.  We will show that under a given set of simple assumptions the
rate of acceleration is actually the same as in the classic case considered
by BR74, i.e., $\Gamma \propto {p_{\rm ext}}^{-1/4}$.

It is unlikely that the magnetic field evolves without some form of
dissipation, especially if it is highly unorganized (reconnection is a
diffusive process, so strong gradients in the field, as are present if the
field is highly tangled on small scales, will likely lead to increased
dissipation).  These loss processes can compete with the efficiency of bulk
acceleration by removing energy from the flow reservoir.  We will
investigate the effects such a tradeoff might have on the dynamics and
appearance of jets.

We will lay out the simplifying assumptions going into this model in
\S\ref{sec:model}.  Section \ref{sec:properties} contains a brief
discussion of allowed solutions and the sonic transition of such solutions.
In \S\ref{sec:selfsimilar} we will discuss self-similar and asymptotic
solutions and present some full analytic solutions, including a simple
estimate of the effects of radiation drag. Section \ref{sec:discussion}
contains a discussion of the aforementioned tradeoff between dissipation
and acceleration, some predicted observational consequences, and an outlook
on applications of this model to radio galaxies and $\gamma$-ray bursts.
Finally, we will summarize our results in \S\ref{sec:conclusions}.


\section{The Model}
\label{sec:model}
The model we are employing here is closely related to (and an extension of)
the `twin-exhaust' model put forward by BR74.  We adopt a similar scenario
under which the jet is launched into a stratified external medium.  In the
case we are considering, turbulent and highly disordered magnetic fields
dominate the internal energetics of the jet.

We have tried to illustrate the overall picture we are employing in Fig.
\ref{fig:Cartoon}: Interstellar magnetic field is advected inward by the
accretion disk.  Turbulent shear then amplifies the field and tangles it up
(dynamo action).  This effect will grow stronger with decreasing distance
to the black hole.  Eventually, regions of very high field strength will
develop.  Due to their buoyancy they will accelerate away from the black
hole, forming an initial outflow.  This outflow is then collimated by the
pressure of the external medium.  The jet channel is constrained by
pressure balance, i.e., the jet will expand or contract in such a way that
an equilibrium solution is set up for which the flow is stationary.  As the
flow expands, we assume that micro instabilities and turbulence constantly
rearrange the field.  As in the pure particle pressure case, the flow can
go through a critical point, where the radius $R$ has a minimum and beyond
which the flow will become self-similar (if the external pressure itself
behaves self-similarly with distance to the black hole) before the rest
mass energy starts dominating the inertia, at which point the jet will
reach a terminal Lorentz factor $\Gamma_{\infty}$.  Along the way, the
field might dissipate energy via reconnection-like processes, and radiation
drag might alter the dynamics.

We assume that no energy or particles are exchanged between jet and
environment, except for radiative losses.  However, the momentum discharge
(the jet thrust $Q$) need not be conserved along the jet.  By assuming that
quantities like $\rho'$, ${U'}_{\rm i}\equiv \langle {{B'}_{\rm
    i}}^2\rangle/8\pi$ (where a prime denotes that the quantity is measured
in the comoving frame) do not vary significantly across the jet we simplify
the analysis to a quasi-1D solution.  We ignore effects of shear at the jet
boundaries.  We assume throughout most of the paper that the advected
matter is cold (i.e., enthalpy density $h' \approx n'm_{\rm particle}
c^2$).  We are looking for stationary flow along the jet (i.e., far from
the terminal shock), enabling us to drop time derivatives.  Finally, to
make this quasi 1D treatment possible, we will need to make the assumption
that the jet is narrow, which in our case implies that the opening angle is
small compared to the beaming angle, i.e., $dR/dz \ll 1/\Gamma$.  As we
will later show, this also implies that the jet is in causal contact with
its environment (as required by the assumption that the jet is in pressure
equilibrium with the surrounding medium).


\subsection{Treatment of Magnetic Field}
\label{sec:magnetic}
We use cylindrical coordinates $(r, \phi, z)$ with the $z$-axis oriented
along the jet axis.  The flow velocity is not aligned with the $z$-axis for
$r>0$ (the jet expands).  We assume that the sideways velocity is small
compared to $v_{\rm z}$ but non-vanishing, i.e., the flow is well
collimated.  The magnetic field is expressed in a different basis, since
the standard basis vectors ${\mathbf e}_{\rm r}$ and ${\mathbf e}_{\rm z}$
are not orthogonal in the comoving frame for $r \not= 0$.  One axis of this
new basis is aligned with the local velocity vector, ${\mathbf
e}_{\parallel}$.  The second unit vector ${\mathbf e}_{\phi}$ is coincident
with the $\phi$-unit vector of the lab coordinate system.  The third unit
vector ${\mathbf e}_{\varpi}$ is obtained from the cross product of the
other two.  In the comoving frame we have
\begin{equation}
  {\mathbf B}' = {B'}_{\varpi}{\mathbf e}_{\varpi} + {B'}_{\phi}{\mathbf
  e}_{\phi} + {B'}_{\parallel}{\mathbf e}_{\parallel}.
  \label{eq:comovingfield}
\end{equation}
As mentioned before, the 1D approximation is only possible if the opening
angle is small compared to the beaming angle.  This is because the Lorentz
factor will not be nearly uniform across the jet otherwise.  The assumption
that $dR/dz \ll 1/\Gamma$ simplifies the equations of motion significantly.

Following B95, who investigated similar jets in the non-relativistic limit,
we assume that the magnetic field is highly disorganized.  In the comoving
frame, averages over the individual components and cross terms vanish while
the energy density in the individual components is not zero:
\begin{eqnarray}
  \langle {B'}_{\rm i}{B'}_{\rm j} \rangle & = & 0,\ {\rm for}\ {\rm
  i}\not={\rm j}, \ \ \ \ \ \ \langle {B'}_{\rm i} \rangle = 0, \nonumber
  \\ \langle {{B'}_{\rm i}}^2 \rangle & \equiv & 8\pi U_{\rm i}' \not=
  0.\label{eq:average}
\end{eqnarray}
Lorentz transformation of the field to the lab frame (and to the
cylindrical coordinate system aligned with the jet axis) yields
\begin{eqnarray}
  {\mathbf B} & = & \left(\frac{v_{\rm z}}{v}\Gamma {B'}_{\varpi} +
  \frac{v_{\rm r}}{v}{B'}_{\parallel}\right){\mathbf \hat{e}}_{\rm r} +
  \Gamma{B'}_{\phi}{\mathbf \hat{e}}_{\phi} \nonumber \\  & & +
  \left(\frac{v_{\rm z}}{v}{B'}_{\parallel} - \frac{v_{\rm r}}{v}\Gamma
  {B'}_{\varpi}\right){\mathbf \hat{e}}_{\rm z} .
\end{eqnarray}
Some of the components are now correlated.  The electric field in the lab
frame is
\begin{equation}
  {\mathbf E} = v_{\rm z}\Gamma {B'}_{\phi}{\mathbf \hat{e}}_{\rm r} -
  \sqrt{{v_{\rm r}}^2 + {v_{\rm z}}^2} \Gamma {B'}_{\varpi}{\mathbf
  \hat{r}}_{\phi} - v_{\rm r}\Gamma{B'}_{\phi}{\mathbf \hat{e}}_{\rm z}.
\end{equation}


\subsubsection{Magnetic Equation of State} \label{sec:mixing} Without the
presence of turbulent rearrangement of the field, flux freezing would
govern the behavior of the individual components.  {If we assume the
presence of turbulent mixing between the different field components, we
might expect the field to follow a modified evolution according to
\begin{eqnarray}
  d{{B'}_{\rm i}} & = & \sum_{\rm j}
    \alpha_{\rm 
    ij} \left. \frac{\partial {{B'}_{\rm j}}}{\partial{\Gamma v}}\right|_{\rm
    ff}d\left(\Gamma v\right) + \sum_{\rm j} \beta_{\rm 
    ij} \left. \frac{\partial {{B'}_{\rm j}}}{\partial{R}}\right|_{\rm
    ff}dR,
\end{eqnarray}
where the subscript ff denotes the value the derivative would take under
flux freezing, and $\alpha_{\rm ij}$ and $\beta_{\rm ij}$ are arbitrary
mixing coefficients.  Based on this picture we therefore choose the
following convenient {\it ad-hoc} parametrization of the field evolution
with Lorentz factor $\Gamma$ and jet radius $R$, including rearrangement:
\begin{eqnarray}
  {{B'}_{\varpi}}^2 & \propto & {{B'}_{\phi}}^2 \ \ \propto \ \
  \left(v\Gamma\right)^{-2+ \mu_{\rm 1}}R^{-2+\mu_{\rm 2}},\nonumber \\
  {{B'}_{\parallel}}^2 & \propto & \left(v\Gamma\right)^{\mu_{\rm
  3}}R^{-4+\mu_{\rm 4}}.
  \label{eq:magstate}
\end{eqnarray}
This is the magnetic equation of state we use.  In the case of pure flux
freezing, $\mu_{\rm i} = 0$ for all ${\rm i}$.  In the case of a completely
isotropic field we have $\mu_{1} = 2/3$, $\mu_{2} = -2/3$, $\mu_{3} =
-4/3$, $\mu = 4/3$.  Note that this prescription is still fully general
[until we make some limiting assumptions about the $\mu_{\rm i} (r,z)$].
Since the re-arrangement process mixes the perpendicular and parallel
components of the field, we would expect that the field behavior is changed
from flux freezing in such a way that the coefficients $\mu_{\rm i}$ are
bracketed by the values they take in the case of flux freezing.  Since the
case of a purely isotropic field must be included in our analysis, it is
clear that this condition requires that $\mu_{2} < 0 < \mu_{1}$ and
$\mu_{3} < 0 < \mu_{2}$.}

We define two quantities to characterize the anisotropy of the magnetic
pressure:
\begin{equation}
  \zeta \equiv \frac{{U'}_{\parallel} - {U'}_{\perp}}{{U'}_{\parallel} +
    {U'}_{\perp}}, \ \ \ \ \ \ \ \ \ \delta \equiv \frac{{U'}_{\phi} -
    {U'}_{\varpi}}{{U'}_{\phi} + {U'}_{\varpi}},
  \label{eq:anisotropies}
\end{equation}
where ${{U'}_{\perp}} \equiv {U'}_{\varpi} + {U'}_{\phi}$.  Thus, the
magnetic field is purely perpendicular for $\zeta = -1$, and purely
parallel for $\zeta = 1$.  The perpendicular component is purely radial for
$\delta = -1$ and purely toroidal for $\delta = 1$.  The field is perfectly
isotropic for $\zeta = -1/3$ and $\delta = 0$.  It is obvious from equation
\,(\ref{eq:magstate}) that $\delta$ is constant for any combination of
parameters, since ${U'}_{\varpi} \propto {U'}_{\phi}$ by assumption.

While this parametrization alone is rather unrestrictive, we can limit it
to a one parameter family by assuming that the $\mu_{\rm i}$ are constants
under {\it any possible} variation of $\Gamma$ and $R$, and that the
rearrangement process does not change the total comoving energy density in
the magnetic field. (Otherwise the same process would have to act as an
energy sink, since we assume that the magnetic field is the dominant term
in the internal energy budget.  We would therefore be dealing with a
dissipative process, which we will address in \S\ref{sec:dissipation}.)  We
can solve for $\mu_{\rm i}$ in terms of $\zeta$ by fixing either $\Gamma$
or $R$ and demanding that the total energy density ${U'} \equiv
\sum{{U'}_{\rm i}}$ behave the same as it would following flux freezing:
\begin{eqnarray}
  d{U'} & = & {U'}_{\perp}\left[\left(\mu_{\rm 1} - 2\right) d(\Gamma v) +
    \left(\mu_{\rm 2} - 2\right) dR \right] \nonumber \\ & & +
    {U'}_{\parallel} \left[\mu_{\rm 3} d(\Gamma v) + \left(\mu_{\rm 4} -
    4\right) dR \right] \nonumber \\ & = & - 2{U'}_{\perp}\left(d(\Gamma v)
    + dR\right) - 4 {U'}_{\parallel} dR
  \label{eq:mixingcondition}
\end{eqnarray}
for arbitrary $d (\Gamma v)$ and $dR$.  Constancy of any of the $\mu_{\rm
  i}$ then implies constancy of $\zeta$ and substitution of $\zeta$ from
  equation \,(\ref{eq:anisotropies}) yields
\begin{eqnarray}
  U_{\rm i} & \propto & \left(\Gamma v\right)^{\zeta - 1} R^{-3 - \zeta}
  \nonumber \\ \mu_{\rm 1} & = & 1 + \zeta, \ \ \ \ \mu_{\rm 2} = -1 -
  \zeta, \nonumber \\ \mu_{\rm 3} & = & \zeta - 1, \ \ \ \ \mu_{\rm 4} = 1
  - \zeta
  \label{eq:mixing}
\end{eqnarray}
which includes the isotropic case, where the magnetic field behaves like a
relativistic gas, for which $\zeta=-1/3$, $\delta=0$.

It turns out that one can find special {analytic} solutions with
constant $\mu_{\rm i}$ that satisfy equation \,(\ref{eq:mixingcondition})
without the requirement that $\zeta$ be constant (see
\S~\ref{sec:selfsimilar}).  For these solutions the rearrangement process
conserves the comoving magnetic energy density {\it only} under the
variations in $\Gamma$ and $R$ allowed by the Bernoulli equation [i.e.,
$d(\Gamma v)$ and $dR$ in equation \,(\ref{eq:mixingcondition}) are not
arbitrary].  The only condition on the parameters $\mu_{\rm i}$ for such a
solution is that $\mu_{\rm 1}/\mu_{\rm 2} = \mu_{\rm 3}/\mu_{\rm 4}$.
These solutions are limited to the self-similar range, where the jet is
dominated by magnetic pressure.  Once they approach the terminal phase
(i.e., $\rho' \gtrsim U'$), the parameters $\mu_{\rm i}$ must vary with
$z$.  For the rest of the paper we will assume that equation
(\ref{eq:mixing}) holds unless indicated otherwise.


\subsubsection{Dissipation of Magnetic Energy}
\label{sec:dissipation}
It is unlikely that the tangled magnetic field evolves without any
dissipation of its energy (e.g., via reconnection).  We thus include a
simple, {\it ad hoc} prescription of magnetic energy losses.  We base our
parametrization on the idea that the magnetic field is always in a nearly
force-free equilibrium.  However, it is impossible to maintain perfect
force-free conditions everywhere and as the jet expands in either
direction, the field responds by rearrangement between the different
components (eq.~[\ref{eq:mixing}]) and by dissipation of some of its
energy.  We therefore assume that the dissipation rate is roughly
proportional to the divergence of the velocity in the comoving frame:
\begin{equation}
  \left.\left(\frac{\partial {U'}_{\rm i}}{\partial
    \tau}\right)\right|_{\rm diss} \approx -\Lambda{U'}_{\rm i}\left|\nabla
    \cdot {\mathbf v}'\right|
\end{equation}
or, in the lab frame
\begin{equation}
  \left.\left(\frac{\partial {U'}_{\rm i}}{\partial z}\right)\right|_{\rm
    diss} \approx -\Lambda{U'}_{\rm i}\left|\frac{\partial}{\partial
    z}\ln{\left(\Gamma v R^2\right)}\right|.
  \label{eq:dissipation}
\end{equation}
This {\it ansatz} can easily be generalized to different $\Lambda_{\rm i}$
for different field components (e.g., if the Alfv\'{e}n velocity factors
into $\Lambda$).  For a more realistic dissipation model see the Appendix.

We will assume that the dissipated energy goes into isotropic particle
pressure, which is then either (a) radiated away immediately as {\it
  isotropic} radiation in the comoving frame or (b) accumulated until
the jet reaches a state of equipartition between particle pressure and
magnetic field.


\subsection{Equations of Motion}
We write the relativistic continuity equation as
\begin{equation}
  \rho' v_{\rm z} \Gamma R^2 = const.
  \label{eq:continuity}
\end{equation}

The energy and momentum equations are given by ${T^{\alpha\beta}}_{;\beta}
= 0$, ($T$ is the stress-energy tensor, separable into a matter and an
electromagnetic part).  In the absence of gravity, this reduces to
${T^{\alpha\beta}}_{,\beta} = 0$, which will be sufficient for the analysis
through most of this paper since most of the acceleration will likely take
place at distance $z \gg r_{\rm g} \equiv GM/c^2$, where $M$ is the mass of
the central black hole.  It turns out, however, that gravity {\it is}
important in discussing the critical points of the jet, in which case we
approximate the covariant derivative by a Newtonian potential
($-{\Gamma^{0}}_{30} = {\Gamma^{3}}_{33} = {\Gamma^{3}}_{00} \approx r_{\rm
  g}/z^2$).  We will comment on the accuracy of this approximation in
\S\ref{sec:sonicpoints}.

Since there is no energy exchange between the jet and the environment, and
using the expression for the electromagnetic field measured in the lab
frame from \S\ref{sec:magnetic}, we can write the energy equation as
${T^{0\alpha}}_{,\alpha} = 0$ (neglecting gravity). We integrate the
equation over a cross-sectional volume of the jet and convert it to a
surface integral using Gauss's law.  The contribution from the sidewall is
zero, giving
\begin{eqnarray}
  \lefteqn{{\Gamma}^2\left(\rho' c^2 + 4p'\right) v_{\rm z} \pi R^2}
    \nonumber \\ & + &
    \frac{1}{4\pi}v_{\rm z}{\Gamma}^2\left({{B'}_{\phi}}^2 +
    {{B'}_{\varpi}}^2\right) \pi R^2 \nonumber \\ & \equiv & L = const.
  \label{eq:energy}
\end{eqnarray}
(where the ${{B'}_{\rm i}}^2$ are now averaged quantities).  Dividing
equation \,(\ref{eq:continuity}) into equation \,(\ref{eq:energy}) gives the
relativistic Bernoulli equation.  In the more general case including
radiative losses and gravity we have
\begin{eqnarray}
  \lefteqn{\frac{d}{dz}\Gamma^2 v_{\rm z} R^2\left({\rho'} c^2 + 4p' +
      2\frac{{{B'}_{\perp}}^2}{8\pi}\right)}   \label{eq:energylosses}
  \\ &  & + 2\frac{r_{\rm g}}{z^2}\Gamma^2 v_{\rm z} R^2 \left(\rho' c^2 +
    4p' + 2{U'}_{\perp}\right) + S_{\rm rad} = 0, \nonumber
\end{eqnarray}
where $S_{\rm rad}$ is the energy lost to radiation leaving the jet.  We
have to make some assumption about the form of $S_{\rm rad}$, i.e., the
amount of energy radiated away [cases (a) and (b) from
\S\ref{sec:dissipation}].

The $z$-momentum flux $Q$ can be calculated in much the same way
(integrating $T^{33}$ across a jet cross-section).  Since the jet can
exchange $z$-momentum with the environment, the momentum discharge need not
be conserved, however.  Dropping terms of order ${v_{\rm r}}^2$, the
integration yields
\begin{eqnarray}
  \lefteqn{Q \equiv \int_{A} d T^{33} \approx \Gamma^2 v^2 \pi R^2(\rho
    + 4p'/c^2) + \pi R^2 p'} \nonumber \\ & & + \pi
  R^2\left[\Gamma^2\left(1
      + v^2\right) {U'}_{\perp} - {U'}_{\parallel}\right].
\end{eqnarray}

The sideways momentum equation is given by ${T^{1\alpha}}_{;\alpha} -
v_{\rm r}{T^{0\alpha}}_{;\alpha} = 0$.  The condition that the solution be
stationary (i.e., $\partial R(z)/\partial t = 0$) gives the pressure
balance condition between the jet and its environment.  We assume that the
internal structure of the field adjusts to maintain the given
cross-section.  Since we {\it assume} that $v_{\rm r} \ll v_{\rm z}$, the
internal variation will be sufficiently small, $\frac{\partial}{\partial r}
\sim \left(\frac{v_{\rm r}}{v_{\rm z}}\right) \frac{\partial}{\partial z}$,
to justify the assumption of uniformity (note: this assumption is only
satisfied if the jet is in causal contact).  We are thus only interested in
the pressure balance condition at the jet walls, $r=R$, which gives
\begin{equation}
  p_{\rm ext} = U_{\phi}' + U_{\parallel}' - U_{\varpi}' + p'.
  \label{eq:pressure}
\end{equation}
Note that $U_{\varpi}'=0$ directly at the jet boundaries, since the
magnetic field is assumed not to penetrate the contact discontinuity.
However, since interior pressure balance demands that $U_{\phi}' +
U_{\parallel}' - U_{\varpi}' + p'$ be constant, we can set $U_{\phi}' -
U_{\varpi}' = const.$ and substitute it for $U_{\phi}'$ at $r=R$, which
gives equation \,(\ref{eq:pressure}).

{\plotfiddle{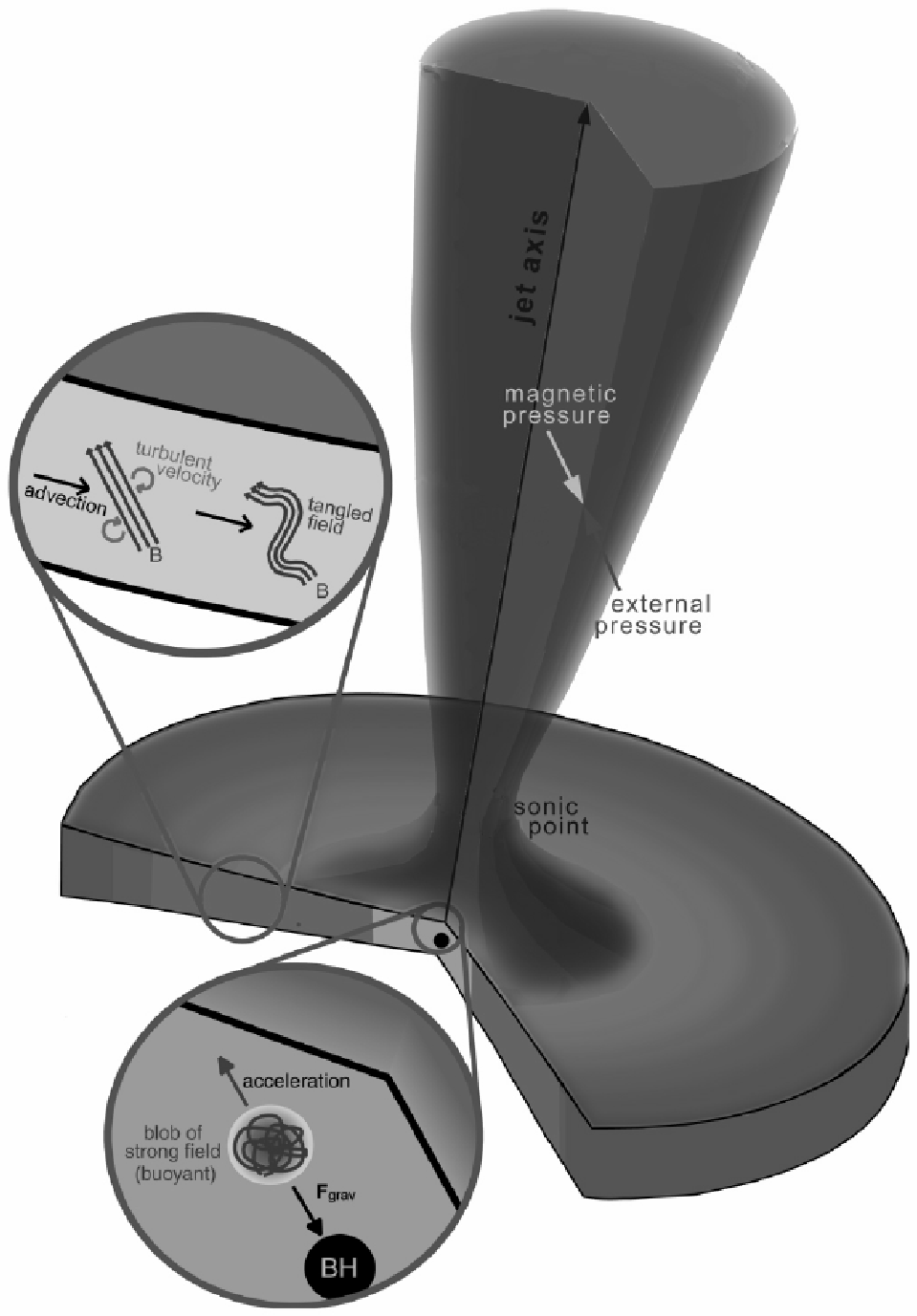}{4.9in}{0}{85}{85}{-261}{-155}
\figcaption{Cartoon of the general picture employed in this article.
  Tangled magnetic field is generated in the disk, advected inward by the
  disk flow, and accelerated away from the black hole to form an initial
  outflow.  Under suitable conditions, this outflow is then collimated and
  accelerates away from the core, keeping pressure balance with the thermal
  pressure provided by the jet environment.\label{fig:Cartoon}}}
\bigskip


\section{Dynamical Solutions}
\label{sec:properties}
Before we start analyzing the equations presented above, it is worth noting
that in the case of a cold ($p'=0$) jet and a magnetic field following pure
flux freezing ($\mu_{\rm 1}=\mu_{\rm 2}=0$) the only possible solution to
equation (\ref{eq:energy}) far away from the core (i.e., $z \gg r_{\rm g}$)
is $\Gamma(z)=const.$, i.e., the jet expands sideways to satisfy pressure
balance, without accelerating.  This is because both the kinetic energy
flux and the Poynting flux do not vary with $R$, while they do vary with
$v$, so that equation (\ref{eq:energy}) becomes an equation of $\Gamma$
only.  Thus, fixing the total jet power $L$ fixes $\Gamma$.  While a
scenario like this might explain the coasting phase of the jet (where no
more acceleration occurs), it cannot account for the initial bulk
acceleration we are looking for.

Note that this is different than the case of anisotropic, relativistic
particle pressure {\it in the absence of isotropization} (i.e., simply
under adiabatic behavior of the individual components).  In that case, the
components scale like $p_{\rm z} \propto (\Gamma v)^{-2}R^{-2}$, $p_{\rm r}
\propto p_{\phi} \propto (\Gamma v)^{-1}R^{-3}$.  We might expect a
behavior like this for a relativistic turbulent pressure term.  The
sideways pressure is simply $p_{\rm r}$.  At relativistic speeds, the
solution approaches the one found by BR74, $\Gamma \propto R \propto
p^{-1/4}$.  Thus, unlike in the magnetic case described in this paper, it
is generally possible to accelerate a jet with anisotropic particle
pressure without making any arbitrary assumptions about the randomization
process.  This is simply because {\it only} the perpendicular component of
the field, ${U'}_{\perp}$, contributes to the Poynting flux, while all
components of the pressure enter equation (\ref{eq:energy}), which
introduces a dependence on $R$, making a solution $\Gamma \not= const.$
possible.  For a magnetically dominated solution to exist, on the other
hand, we need a field rearrangement process at work, such as was described
in \S\ref{sec:mixing}.  But even under such favorable conditions, a proper,
accelerating solution is not always guaranteed.


\subsection{Critical Points}
\label{sec:sonicpoints}
Since the jet will likely be injected with sub-relativistic speed, the
question arises as to where the jet crosses possible critical points and at
what velocity.  If the jet is injected at large distances from the central
black hole, we can neglect gravity; if it is injected close to the hole we
will have to include at least a phenomenological gravity term.

As a first step, we will set $M = 0$ artificially (still assuming the
presence of an external pressure gradient) and neglect dissipation ($S_{\rm
  rad} = 0$).  Equation (\ref{eq:energylosses}) gives
\begin{eqnarray}
  \lefteqn{{\Gamma}^2\left[\left(\rho'c^2 + 4p' +
    2{U'}_{\perp}\right)v^2\right.}
    \nonumber \\
    & + & \left.4\left(1 - \gamma_{\rm ad}\right) p' + 2\zeta
    {U'}_{\perp}\right]\frac{dv}{vdz} \nonumber \\
    & + &  \left\{4\left(2 - \gamma_{\rm ad}\right)p' - 2\left(1 +
      \zeta\right){U'}_{\perp}\right\}\frac{dR}{Rdz} = 0,
  \label{eq:soundequation}
\end{eqnarray}
where $\gamma_{\rm ad}$ is the adiabatic index of the particles,
$\gamma_{\rm ad} \equiv d\ln{p'}/d\ln{\rho'}$.  This equation has a
critical point when the expression in square brackets vanishes.  At such a
point, the jet cross-section must satisfy $dR/dz=0$, i.e., the jet must go
through a nozzle, the position and cross section of which are determined by
the dynamics of the flow.  Following the notation of BR74, the velocity at
which that happens is
\begin{equation}
  c_{\star} = \sqrt{\frac{4{p'}_{\star}\left(\gamma_{\rm ad} -1\right) -
      2{{U'}_{\perp}}_{\star}\zeta}{{\rho'}_{\star}
      c^2 + 4{p'}_{\star} + 2{{U'}_{\perp}}_{\star}}},
  \label{eq:soundspeed}
\end{equation}
where the subscript ${\star}$ indicates that the quantity is evaluated at
the critical point $z_{\star}$.  Since for a magnetically dominated jet
$\lim_{\zeta \rightarrow 0-} c_{\star} = 0$, the critical point exists only
for $\zeta < 0$.  For a purely isotropic field, where $\zeta = -1/3$,
$c_{\star}$ reduces to the sound speed of a relativistic gas with
$\gamma_{\rm ad} = 4/3$, $c_{\star} = \sqrt{1/3}$.

Locally we can always write $p_{\rm ext} \propto z^{-\xi}$, thus we define
\begin{equation}
  \xi \equiv -d\ln{p_{\rm ext}}/d\ln{z}.
  \label{eq:xi}
\end{equation}
We can then substitute the pressure balance condition (\ref{eq:pressure})
into equation (\ref{eq:soundequation}) in the limit $\rho'c^2 + 4p' \ll
{U'}$ and eliminate $R$, which yields
\begin{equation}
  {\Gamma}^2\left[\left(3 + \zeta\right)v^2 + \left(1 +
      3\zeta\right)\right]\frac{dv}{vdz} = \left(1 +
    \zeta\right)\frac{\xi}{z}
  \label{eq:criticalequation}.
\end{equation}
This equation also has a critical point with a critical speed of
\begin{equation}
  c_{\dagger} \equiv \sqrt{\frac{1 + 3\zeta}{-3 - \zeta}}.
  \label{eq:criticalspeed}
\end{equation}
Unlike equation \,(\ref{eq:soundequation}), solutions cannot cross this critical
point, since there $dv/dz \rightarrow \infty$ (but see
\S\ref{sec:gravity}).

We expect $dp_{\rm ext}/dz < 0$, so solutions always accelerate
(decelerate) for $v > c_{\dagger}$ ($v < c_{\dagger}$).  Since
$c_{\dagger}$ only exists for $\zeta < -1/3$, solutions with $\zeta > -1/3$
always accelerate.  In that case equation \,(\ref{eq:soundequation}) implies
that for $v > c_{\star}$ ($v < c_{\star}$) the jet is expanding
(contracting) in the $r$-direction. Since $c_{\star}$ only exists for
$\zeta < 0$, solutions with $\zeta > 0$ always expand sideways.

If, on the other hand, $\zeta < -1/3$, two branches of solutions exist: (a)
solutions which are injected with $v > c_{\dagger}$, which always
accelerate and go through a nozzle at $v = c_{\star} \geq c_{\dagger}$, and
(b) solutions which are injected with $v < c_{\dagger}$, which always
decelerate.  Thus, at sufficiently large distances from the core for
gravity to be negligible (see \S\ref{sec:gravity}), highly anisotropic
solutions with $\zeta \approx -1$ have to be injected at relativistic
velocities to be accelerating, since $c_{\dagger} \rightarrow 1$ as $\zeta
\rightarrow -1$.  This corresponds to the right branch of the dashed
solutions plotted in Fig.~\ref{fig:sonic} (which includes the effects of
gravity, see \S\ref{sec:gravity}).

It is instructive to look at the case of pure {\it anisotropic}
relativistic particle pressure again.  We define the pressure anisotropy as
$\zeta_p \equiv (p_{\parallel} - p_{\perp})/(p_{\perp} + p_{\parallel})$.
If we fix $\zeta_p$ by some rearrangement process as we did for the
magnetic field in \S~\ref{sec:mixing} (which might occur, for example, if
there is coupling between magnetic field and turbulent pressure as
suggested by B95), the behavior is very similar in the sense that
accelerating solutions for $\zeta_p > -1/3$ (i.e., for $p_{\perp} >
2p_{\parallel}$) have to be injected at super-critical velocity $v >
c_{\dagger_p} = \sqrt{(1+3\zeta_p)/(5-\zeta_p)}$.

If we simply let the components of $p$ evolve adiabatically (without
rearrangement), we arrive at a different critical velocity, $c_{\dagger_p}
= \sqrt{(5 + 7\zeta_p)/(9 + 3\zeta_p)}$ (note that $\zeta$ is not constant
in this case), which exists only for $\zeta_p > - 5/7$. The solution once
again has a nozzle at $c_{\star_p} = \sqrt{(2 + 2\zeta_p)/(3 + \zeta_p)}$.
Since $c_{\star_p} > c_{\dagger_p}$ for $\zeta_p > -5/7$, and since $dR/dz
< 0$ for $v < c_{\star_p}$, solutions injected with $v > c_{\dagger_p}$
must accelerate to satisfy pressure balance, which means that $\zeta_p$
increases with $z$, reducing $c_{\dagger_p}$ and thus making the flow {\it
  more} super-critical (i.e., once above the critical point, the solution
moves away from it).


\subsection{The Effects of Gravity on the Sonic Transition}
\label{sec:gravity}
As seen in the previous section, a solution for $\zeta > -1/3$ that starts
out with $v < c_{\dagger}$ will always decelerate in the absence of
gravity.  However, as is the case in the solar wind, gravity can actually
help a flow go through a critical point.  We thus consider $M > 0$ in this
section.

We can go through the same arguments as in \S\ref{sec:sonicpoints}.  The
critical speeds are still given by equations (\ref{eq:soundspeed}) and
(\ref{eq:criticalspeed}), but now the critical conditions are different.
At $c_{\star}$, equation \,(\ref{eq:energylosses}) gives
\begin{eqnarray}
  \lefteqn{\left[4{p'}_{\star} \left(2 - 2\gamma_{\rm ad}\right) -
      2\left(1 +
        \zeta\right) {{U'}_{\perp}}_{\star}\right]\frac{dR}{R dz}}
  \nonumber \\
  & & + \left({\rho'}_{\star} c^2 + 4{p'}_{\star} +
    2{{U'}_{\perp}}_{\star}\right) \frac{2r_{\rm g}}{z_{\star}}^2 = 0
\end{eqnarray}
instead of $dR/dz = 0$.  Since $\zeta \geq -1$ and $\gamma_{\rm ad} > 1$,
we can infer that $dR/dz > 0$ at $z_{\star}$, i.e., there is no `geometric'
nozzle at $z_{\star}$ anymore.  The solution can always adjust $dR/dz$ to
satisfy this condition, thus the first critical point $c_{\star}$ becomes
irrelevant.

Inclusion of the gravity term changes equation
\,(\ref{eq:criticalequation}) to
\begin{eqnarray}
  \lefteqn{{\Gamma}^2\left[\left(3 + \zeta\right)v^2 + \left(1 +
        3\zeta\right)\right]\frac{dv}{vdz}} \nonumber \\
  & = & \left(1 +
    \zeta\right)\frac{\xi}{z} - \left(3 + \zeta\right)\frac{2r_{\rm
      g}}{z^2}.
  \label{eq:criticalequation2}
\end{eqnarray}

Now solutions can cross the critical point $c_{\dagger}$, since the right
hand side of equation \,(\ref{eq:criticalequation2}) vanishes at
\begin{equation}
  z_{\dagger} \equiv \left(\frac{3 + \zeta}{1 +
\zeta}\right)\frac{2r_{\rm
  g}}{\xi}.
  \label{eq:zcritical}
\end{equation}

If $v \not= c_{\dagger}$ at $z_{\dagger}$, the solution must follow $dv/dz
= 0$ at that point.  This is true for {\it all} $\zeta$.  Since that is the
only zero of equation \,(\ref{eq:criticalequation2}), we can therefore
conclude that solutions accelerating at any $z > z_{\dagger}$ will
accelerate for {\it all} $z > z_{\dagger}$.

Solving the equations for $dR/dz$ instead gives
\begin{eqnarray}
  \lefteqn{\left[\left(\zeta^2 - 1\right) - \left(3 +
        \zeta\right)\left(v^2 +
        \zeta\right)\right]\frac{dR}{Rdz}} \nonumber \\
  & = & -\left(v^2 +
    \zeta\right)\frac{\xi}{z} + \left(\zeta - 1\right)\frac{2r_{\rm
      g}}{z^2},
  \label{eq:radcriteq}
\end{eqnarray}
which also has a critical point at $c_{\dagger}$.  For $v=c_{\dagger}$, the
right hand side of this equation {\it only} vanishes at $z_{\dagger}$.  In
that case, $dR/dz$ remains finite.  For all other solutions (i.e., if $z
\not= z_{\dagger}$ when $v=c_{\dagger}$), we must have singularities in
both $dv/dz$ and $dR/dz$.  The singularity in $dv/dz$ is evident from
Fig.~\ref{fig:sonic} and from equation \,(\ref{eq:criticalequation2});
pressure balance then requires that $dR/dz$ must have a singularity of
opposite sign, since $dp_{\rm ext}/{dz}$ is assumed to be finite, i.e.
$p_{\rm ext}$ is continuous.

We have numerically integrated equation \,(\ref{eq:criticalequation2}) for
two representative cases ($\zeta = -0.9$ and $\zeta = 0$, $\xi = 2$) and
plotted them in Fig.~\ref{fig:sonic}.  Solutions are qualitatively
different for $\zeta < -1/3$ and $\zeta > -1/3$:
\begin{itemize}
\item{For $\zeta < -1/3$, there is {\it one} accelerating transonic
    solution, given by the condition in equation \,(\ref{eq:zcritical}),
    shown in the upper panel of Fig.~\ref{fig:sonic} as a thick solid black
    curve.  This is also the only solution accelerating for all $z$.  As in
    the case of a regular adiabatic wind (Parker 1958), there also exists a
    decelerating transonic solution.  Regions where solutions contract in
    the $r$-direction (i.e., where $dR/dz < 0$) are shown as hatched areas.
    There are four more branches of solutions. Two branches are
    double-valued (shown as dashed curves in Fig.~\ref{fig:sonic}).  The
    left branch of those solutions can be rejected since those solutions
    only exist for $z < z_{\dagger}$.  For an accelerating solution on the
    right branch to exist, it must be injected with $v > c_{\dagger}$.
    This corresponds to the solutions discussed in \S\ref{sec:sonicpoints}
    for which gravity can be neglected.
    
    The remaining two branches are solutions that are always sub- or
    supercritical (plotted as thin solid black curves in
    Fig.~\ref{fig:sonic}).  The sub-critical solutions decelerate for large
    $z$ and always stay sub-relativistic.  They are uninteresting as
    possible candidates for relativistic jets.  The supersonic solutions
    decelerate for $z < z_{\dagger}$ and accelerate for $z > z_{\dagger}$.
    These solutions correspond to the super-critical solutions mentioned in
    \S\ref{sec:sonicpoints}.  They always expand in the sideways direction.
    As we let $\zeta \rightarrow -1$, $z_{\dagger} \rightarrow \infty$.
    This is not necessarily an indication that no solution is possible for
    $\zeta \approx -1$, since for those cases the critical speed is very
    close to 1, thus the solution can attain large $\Gamma$.  Furthermore,
    as we saw above, the solution is expanding even before it goes through
    $z_{\dagger}$. We can thus have a regular (though sub-critical)
    accelerating jet even for $\zeta \approx -1$.}
\item{For $\zeta > -1/3$, there is only one branch of solutions, all of
    which start out decelerating, shown in the bottom panel of
    Fig.~\ref{fig:sonic}.  As the solutions reach $z_{\dagger}$ they begin
    to accelerate and behave the same way as described in
    \S\ref{sec:sonicpoints}.  Since $z_{\dagger}$ moves inward for
    increasing $\zeta$, this is no handicap.  For $\zeta > -1/3$ we have
    $z_{\dagger} < 8 r_{\rm g}/\xi$ from equation \,(\ref{eq:zcritical}),
    which, for reasonable values of $\xi$, is well in the regime where
    relativistic corrections become important and inside the region where
    we expect the injection to occur.  All of these solutions have positive
    sideways expansion $dR/dz > 0$ for all $z$.}
\end{itemize}

The transonic solution for $\zeta < -1/3$ has some additionally nice
features: Since we know $z_{\dagger}$, we can relate the jet cross section
to the total jet power $L_{\dagger}$ at $z_{\dagger}$.  Assuming that the
jet is still magnetically dominated at $z_{\dagger}$, the kinetic
luminosity of the jet is
\begin{equation}
  L_{\dagger} = \pi {R_{\dagger}}^2 {\Gamma_{\dagger}}^2{v_{\dagger}}
  2{{U'}_{\perp}}_{\dagger}.
\end{equation}
The external pressure at $z_{\dagger}$ is ${p_{\rm ext}}_{\dagger}$ and so
\begin{equation}
  R_{\dagger} = \sqrt{\frac{L_{\dagger} \left(\delta + \frac{1 +
\zeta}{1 -
          \zeta}\right)\left(1+ \zeta\right)}{\pi {p_{\rm
ext}}_{\dagger}
      \sqrt{-\zeta}}}.
  \label{eq:power}
\end{equation}

While we do not know ${p_{\rm ext}}_{\dagger}$, for most parameter choices
$L_{\dagger}$ is very nearly equal to $L_{\infty}$, which we have a
reasonably good handle on from an observational point of view.
Furthermore, we can estimate the jet width at observable distances and
scale the solution back to $z_{\dagger}$, which gives us an estimate of
$p_{\dagger}$ and thus ${{U'}_{\perp}}_{\dagger}$.  This in turn will allow
us to determine the original matter loading of the jet from estimates of
the terminal Lorentz factor $\Gamma_{\infty}$.

{\plotfiddle{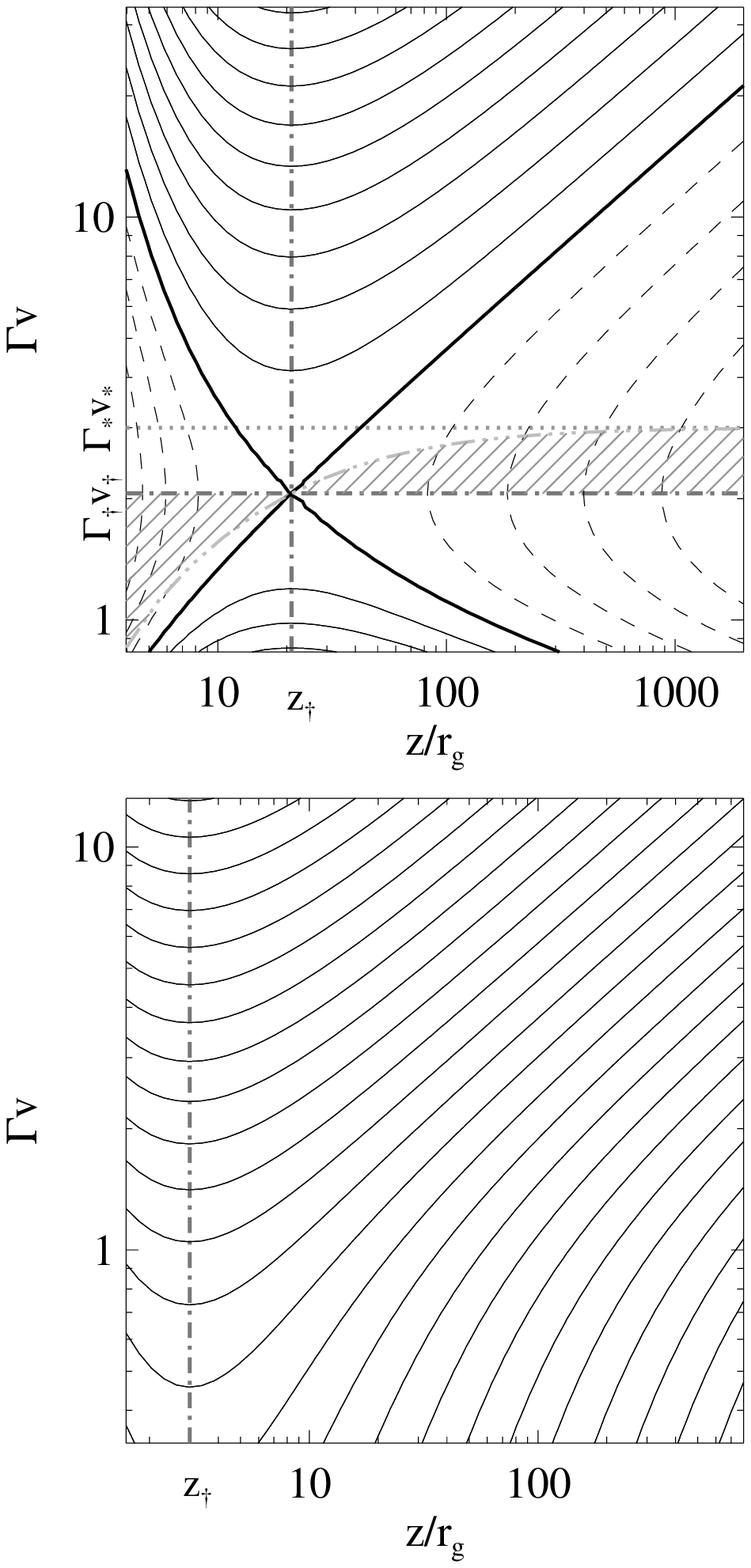}{7.5in}{0}{90}{90}{-191}{0}
\figcaption{Inner solutions for $\xi = 2$ in the vicinity of the critical
  point $z_{\dagger}$ in the absence of dissipation but including
  gravity for the two possible cases, $\zeta < -1/3$ and $\zeta >
  -1/3$.  In the first case (panel a), shown for $\zeta = -0.9$, the
  transonic solutions are thick black lines, sub- and super-critical
  solutions are thin black lines.  Dashed lines show double-valued
  solutions (of interest is only the upper right quadrant for
  $z>z_{\dagger}$, $v>c_{\dagger}$).  The critical values
  $c_{\dagger}$, $z_{\dagger}$ are shown as dash-dotted lines,
  $c_{\star}$ is plotted as a grey dotted line.  The condition
  $dR/dz=0$ is shown as a dash-tripple-dotted line and regions where
  $dR/dz < 0$ are shown as hatched areas.  In the second case (panel
  b, for $\zeta = 0$) only super-critical solutions are possible.  All
  solutions are expanding sideways and all possible initial values
  lead to acceptable solutions.  We have once again indicated the
  location of $z_{\dagger}$.\label{fig:sonic}}}
\bigskip


\section{Solutions in the Self-Similar Regime and Asymptotic Solutions}
\label{sec:selfsimilar}
For an already relativistic jet in the `self-similar' range $z_{\dagger}
\ll z \ll z_{\rm inertia}$ (where $z_{\rm inertia}$ is the location where
$\rho' c^2 = 2{U'}_{\perp}$) a self-similar solution can be found.
Equations\,(\ref{eq:mixing}) and (\ref{eq:energy}) give $\Gamma \propto
R^{-\mu_{\rm 2}/\mu_{\rm 1}}$ and with equation \,(\ref{eq:mixing}) we have
$\Gamma \propto R$.

Under the conditions of equation \,(\ref{eq:mixing}), the pressure balance
condition gives
\begin{equation}
  \Gamma \propto R \propto {p_{\rm ext}}^{-1/4},
\end{equation}
the same as in the case of isotropic particle pressure considered by BR74.
For future reference we define the acceleration efficiency
\begin{equation}
  \eta \equiv -\frac{d\ln{\Gamma}}{d\ln{p_{\rm ext}}},
\end{equation}
thus for this simple case $\eta = 1/4$.

If we adopt the less limiting restriction $\mu_{\rm 1}/\mu_{\rm 2} =
\mu_{\rm 3}/\mu_{\rm 4}$ (see \S\ref{sec:mixing}) instead of equation
\,(\ref{eq:mixing}), we can find powerlaw solutions in three limiting
cases:
\begin{itemize}
\item[(a)]{For $\delta = 0$, the solution is given by $\Gamma \propto
    R^{-\mu_{\rm 2}/\mu_{\rm 1}} \propto {p_{\rm ext}}^{\mu_{\rm
    2}/4\mu_{\rm 1}}$, which can be very efficient for $\mu_{\rm 1}
    \ll -\mu_{\rm 2}$ {(as pointed out in \S\ref{sec:mixing}, one
    would generally expect that $\mu_{\rm 2} < 0 < \mu_{\rm 1}$.}}
\item[(b)]{For $\zeta \approx -1$, the solution is given by $\Gamma \propto
    R^{-\mu_{\rm 2}/\mu_{\rm 1}} \propto {p_{\rm ext}}^{1/(2\mu_{\rm
        1}/\mu_{\rm 2} - 2)}$, which has a limiting efficiency of $\eta
    \leq 1/2$.}
\item[(c)]{For $\zeta \approx 1$ the solution is approximately the same as
    case (a).}
\end{itemize}
If $\delta > 0$, the solution will in general approach solutions (b) or (c)
(i.e., $\zeta \rightarrow \pm 1$).  For $\delta < 0$ it is possible that
the solution approaches a finite terminal Lorentz factor and zero opening
angle if radial tension cancels the pressure due to $U_{\parallel}$ and
$U_{\phi}$.  Fig. \ref{fig:parameterspace} shows the different regimes.
Note once again that these solutions exist only for $\Gamma \gg 1$ and
$\rho c^2 \ll U_{\perp}$. For all other cases the coefficients $\mu_{\rm
  1}$, $\mu_{\rm 2}$, $\mu_{\rm 3}$, and $\mu_{\rm 4}$ are not constant.
{Note that we would generally expect that $\mu_{1} \sim -\mu_{2}$,
  since otherwise the re-arrangement process would be acting
  preferentially for changes in geometry in one specific direction, which
  seems arbitrary. Thus, these results reduce to the well known $\eta \sim
  1/4$.}

We can look for solutions in the presence of dissipation of magnetic
energy.  We now have to consider equation \,(\ref{eq:energylosses}).  We
assume that the energy goes completely into relativistic particles, thus
energy conservation implies
\begin{equation}
  \left.\frac{dp'}{dz}\right|_{\rm dissipation} =
-\left.\frac{1}{3}\frac{dU'}{dz}\right|_{\rm dissipation}.
\end{equation}
The particle energy can subsequently be radiated away as isotropic
radiation.  As long as $p' \ll {U'}_{\perp}$, the radiative case is no
different from the non-radiative case, since the adiabatic term in the
particle pressure does not contribute to the dynamics.

{\plotfiddle{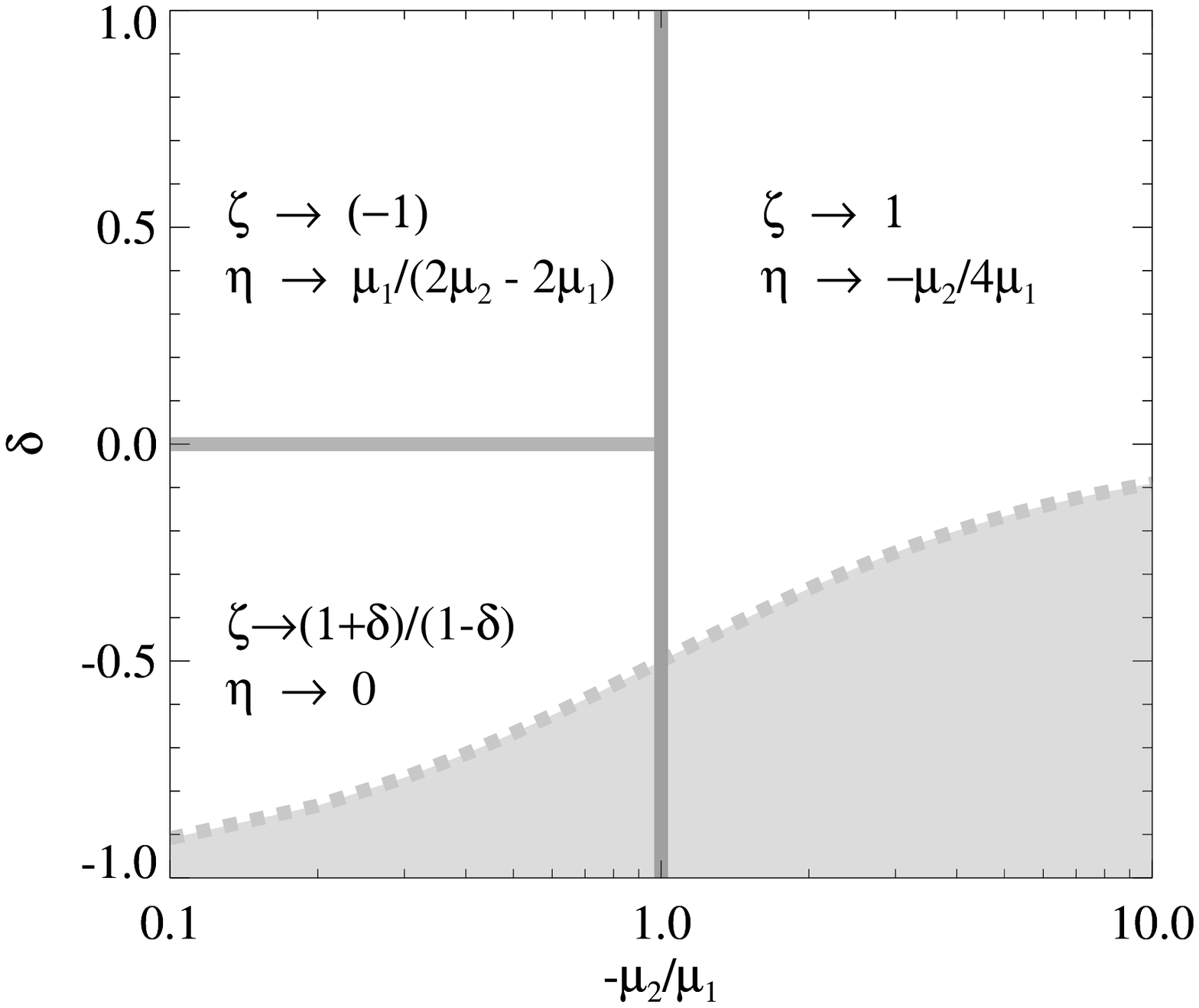}{3.1in}{0}{50}{50}{-128}{0}
\figcaption{Regions in parameter space for the most general choice of
  parameters $\mu_{\rm i}$ possible in the self-similar region.  We
  have indicated the various limits of $\zeta$ at sufficiently low
  external pressure.  Along with $\zeta_{\infty}$ we have indicated
  the limiting efficiency achievable when $\zeta$ has reached is
  limiting value.  For $-\mu_{\rm 2}/\mu_{\rm 1} > 1$, the efficiency
  $\eta$ can become very large.  For $\delta < 0$, it is possible that
  the jet stalls due to radial tension. Note also that for $\delta <
  0$ there is a minimum value for $\zeta$ at the injection below which
  no solution is possible, given by $\zeta_{\rm min} >
  (1+\delta)/(\delta-1)$.  Even if $\zeta_0 > \zeta_{\rm min}$, the
  solution will contract and decelerate if $\delta$ is below a certain
  value.  We have plotted this parameter range for a jet that starts
  out isotropically (i.e., with $\zeta=-1/3$) as a grey area.  For
  smaller $\zeta$, this area will become
  larger.\label{fig:parameterspace}}}
\bigskip

A powerlaw type solution is once again possible only if the enthalpy is
negligible compared to the magnetic energy density (see
\S\ref{sec:equipartition}).  In that case the solution in the self-similar
range is given by $\Gamma \propto {p_{\rm ext}}^{-\eta}$ with $\eta = {1/[4
  + 6\Lambda(3 + 5\zeta)/(3 - 3\zeta^2 - 2\Lambda - 6\Lambda\zeta)]} \leq
1$, with $\eta = 1$ at $\zeta = -1$.  We have plotted $\eta$ as a function
of $\zeta$ and $\Lambda$ in Fig.  \ref{fig:IsoDiss}.  For $\zeta < -3/5$
the efficiency is larger than in the case without dissipation, for $\zeta >
-3/5$ it is smaller.  The limiting efficiency that can be achieved in such
a jet under the condition that $dR/dz > 0$ is given by $\eta \leq 1/2$, as
$\Lambda \rightarrow 0$.  This happens because the dissipative process can
convert energy in the parallel field component $U_{\parallel}$ (which does
not enter eq.~[\ref{eq:energy}]) into particle pressure, which must be
taken into account in the energy balance.  Also shown are areas in
$\zeta$-$\Lambda$-space where the conditions $d\Gamma/dz > 0$, $dR/dz > 0$
are not satisfied.


\subsection{Opening Angles and Causal Contact}
\label{sec:openingangle}
Since a jet is generally defined as a collimated outflow, we can ask under
which conditions the solutions from above are actually collimated.  The
collimation condition $dR/dz < 1$ translates to a pressure gradient
$-d\ln{p_{\rm ext}}/d\ln{z} = \xi < 4$ as long as the jet is magnetically
dominated, the same as in the particle pressure dominated case.  The
presence of dissipation can alter this value.  Generally, the collimation
is increased by dissipation, since the sideways pressure is reduced, thus
the jet does not need to expand as much.  In the coasting phase, where the
jet is no longer accelerating, this condition changes to $\xi < 3 + \zeta <
4$.

Given the solutions from above, we can investigate the ratio of the opening
angle $\alpha_{\rm o}$ to the beaming angle $\alpha_{\rm b} \sim
\Gamma^{-1}$.  In the absence of dissipation we can write
\begin{equation}
  \alpha_{\rm o}/\alpha_{\rm b} = \Gamma\frac{dR}{dz} \propto z^{\xi/2 -
    1},
\end{equation}
independent of $\zeta$.  Thus, for steep pressure gradients $\xi > 2$, the
opening angle will grow faster than the beaming angle and will thus always
become larger even if it starts out being smaller.  For shallow pressure
gradients $\xi<2$, the situation is reversed, i.e., the beaming angle will
eventually become larger than the opening angle.  The presence of
dissipation changes this behavior qualitatively: the ratio $\alpha_{\rm
  o}/\alpha_{\rm b}$ now depends on both $\Lambda$ and $\zeta$, as
illustrated in Fig. \ref{fig:IsoDiss}.

{\plotfiddle{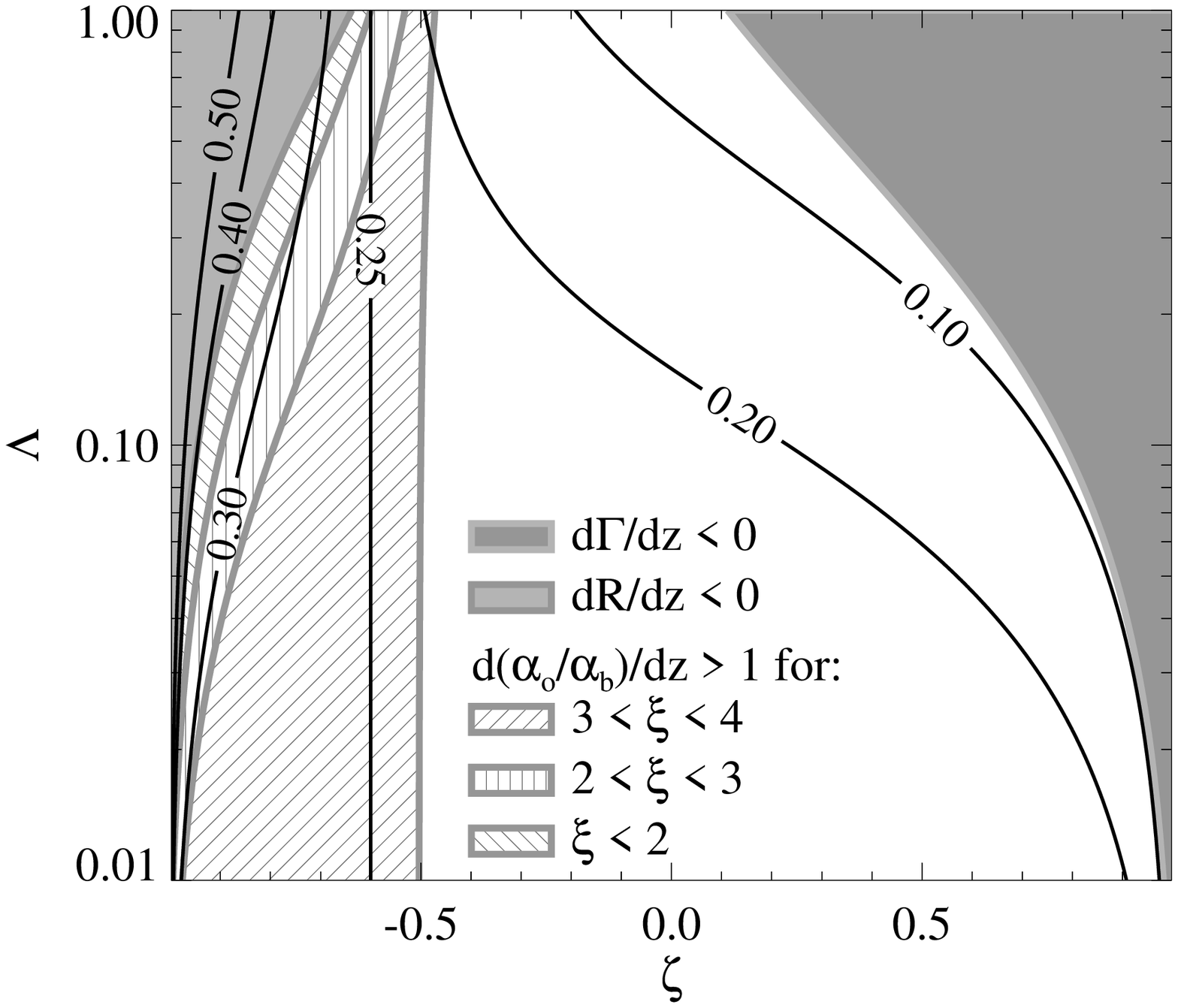}{3.1in}{0}{50}{50}{-128}{0}
\figcaption{Contour plot of acceleration efficiency $\eta \equiv -
  d\ln{\Gamma}/d\ln{p_{\rm ext}}$ for radiative dissipation and
  different values of $\zeta$ and $\Lambda$.  The grey areas indicate
  parameter values for which the jet is either not collimated (light grey
  with dark grey border) or decelerates (dark grey with light grey border).
  The hatched areas indicate regions where the ratio $\alpha_{\rm
  o}/\alpha_{\rm b}$ of opening angle to beaming angle grows with $z$ for
  given values of $0 < \xi < 4$.\label{fig:IsoDiss}}}
\bigskip

This has consequences for the appearance of the jet, since the effective
beaming angle is given by the larger of the two angles.  Under the
assumption that the jet is always collimated, the opening angle in the
coasting phase will always become smaller than the beaming angle, since the
jet does not accelerate anymore.  This could have important consequences
for the morphology of superluminal sources: if the beaming angle were
smaller than the opening angle, one might expect to see larger jet
misalignments, or lose the jet morphology altogether.  The appearance would
become sensitive to the emissivity and local Lorentz factor as a function
of position across the jet cross section.

Jets that expand too fast will eventually lose causal contact with their
environment.  This happens when the Alfv$\acute{\rm e}$n crossing time of
the jet becomes larger than the expansion time (in the comoving frame),
i.e., when
\begin{equation}
  {\tau'}_{\rm A} = \frac{R}{v_{\rm Alfv\acute{e}n}} \approx \frac{R}{c} >
  {\tau'}_{\exp} = \frac{p'}{\Gamma v d{p'}/dz} \approx \frac{z}{\Gamma c
  \xi}
\end{equation}
(where we approximated $v_{\rm Alfv\acute{e}n} \sim c$) or
\begin{equation}
  R > \frac{z}{\Gamma \xi}.
\end{equation}
This corresponds (up to the factor $\xi$) to the criterion when
$\alpha_{\rm o} > \alpha_{\rm b}$.  Thus, for $\xi > 2$ (in the absence of
dissipation) the jet will eventually lose causal contact with its
surroundings (see Fig. \ref{fig:IsoDiss} for values of $\zeta$ and $\xi$
where this is the case).  As mentioned in \S\ref{sec:model}, a quasi 1D
treatment is no longer possible, since the internal pressure balance is now
regulated by shocks traveling inward from the jet walls.  After the jet
reaches the terminal phase, it will re-gain causal contact, since the
opening angle will continually decrease (assuming the jet is still
collimated).


\subsection{Equipartition}
\label{sec:equipartition}
Constant pumping of magnetic energy into particle pressure can lead to
equipartition between $U'$ and $p$.  We can use the self-similar solution
to estimate $\Gamma_{\rm eq}$, where the accumulated particle pressure
surpasses the magnetic energy density (including effects of adiabatic
cooling on the accumulated particle pressure, where we assume that it
behaves as a relativistic gas, which gives an upper limit on the resulting
pressure).  Thus, for $\Gamma_{\infty} \gtrsim \Gamma_{\rm eq}$ the
solution might be altered.  For some parameter values $p$ never reaches the
level of ${U'}_{\perp}$.  In that case we estimate the asymptotic ratio
$\left(p'/{U'}_{\perp}\right)_{\infty}$.  Figure \ref{fig:GammaEq} shows
the results of those estimates.  For large $\zeta$, equipartition can be
reached quickly, thus, unless the energy going into particles is
subsequently radiated away, the assumption that the particle pressure be
negligible compared to the magnetic field energy density will be violated
beyond $\Gamma_{\rm eq}$.

{\plotfiddle{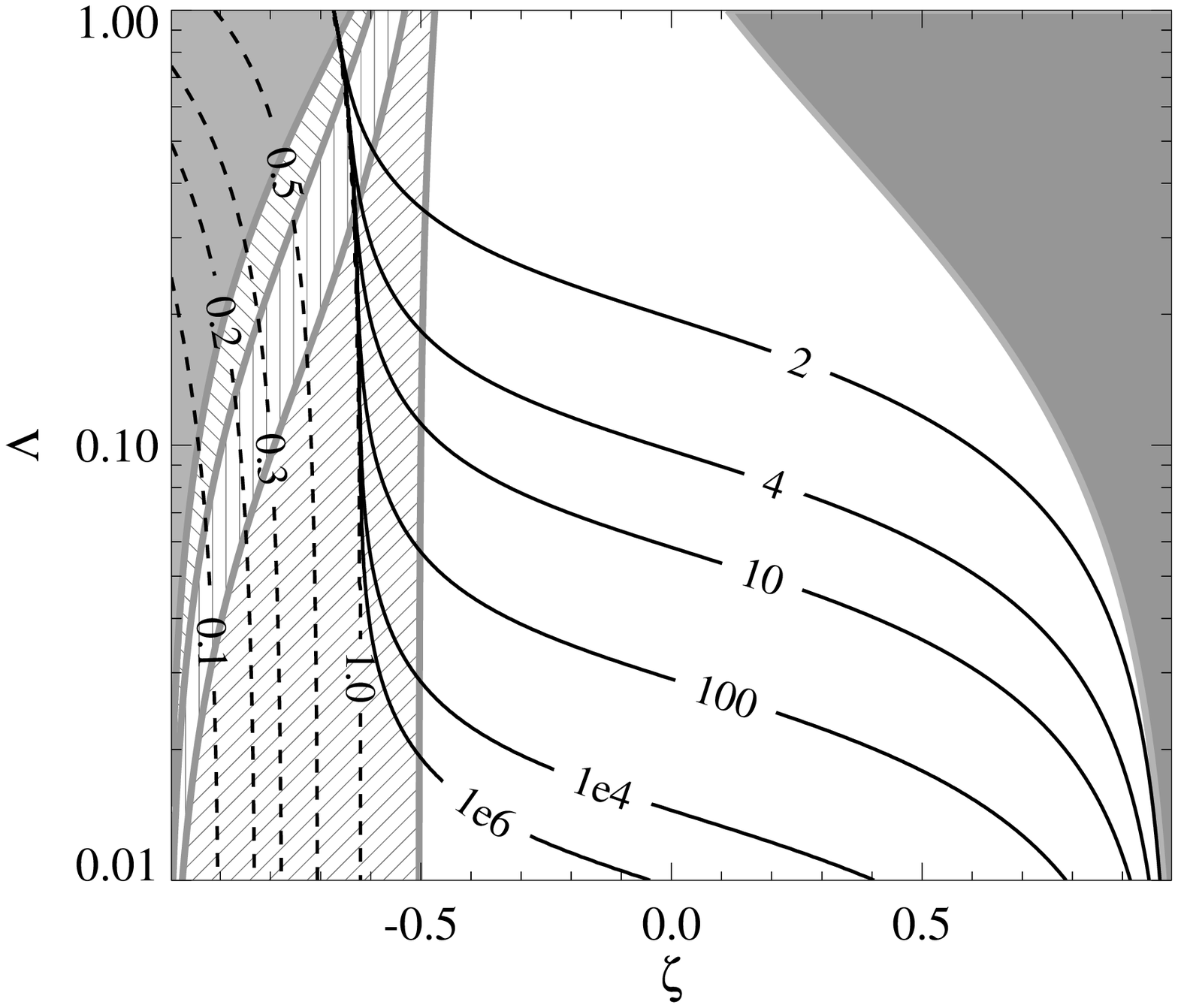}{3.0in}{0}{50}{50}{-128}{0}
\figcaption{Value of $\Gamma$ for which the pressure accumulated by
  non-radiative dissipation reaches equipartition with ${U'}_{\perp}$.
  Neglecting $p$ is no longer justified beyond that $\Gamma$.  For $\zeta
  \lesssim -0.6$ the pressure never catches up with ${U'}_{\perp}$, in that
  case we plotted contours of the limiting ratio $p'/{U'}_{\perp}$ (dashed
  lines).  The hatched and grey regions correspond to the regions in Fig.
  \ref{fig:IsoDiss}.\label{fig:GammaEq}}}
\bigskip

We define the energy distribution function as
\begin{equation}
  f(\gamma) \equiv F_0 \gamma^{-s},\ \ \ \ \ \int_{\gamma_1}^{\gamma_2}
  f(\gamma) d\gamma = {n'},
  \label{eq:spectrum}
\end{equation}
where $\gamma$ is the Lorentz factor according to a particle's random
motion, measured in the comoving frame, and $\gamma_1 \ll \gamma_2$ are the
lower and upper spectral cutoffs.  Since we assume that the magnetic field
is dominating the internal energy budget, synchrotron losses can be very
strong, provided the particle energy spectrum is flat enough so that most
of the energy is at high particle energies (i.e., $s < 2$).  In that case
we can expect most of the energy to be radiated away immediately and the
corresponding electrons will lose most of the inertia, thus the dissipated
energy will not lead to a build-up of particle pressure.  If, however,
synchrotron losses are weak compared to dissipation (e.g., if the spectrum
is too steep, or if synchrotron self-absorption traps most of the radiation
to inhibit cooling), the effects of particle pressure can become important,
as demonstrated in Fig.~\ref{fig:GammaEq}.  For a discussion of the
observational effects of the different radiative scenarios see
\S\ref{sec:tradeoff}


\subsection{Full Solutions}
\label{sec:solutions}

We can solve the full equation (\ref{eq:energylosses}) in the regime
$\Gamma \gg 1$, i.e., for relativistic jets.  As mentioned before, the
pressure balance condition leads to a simple algebraic equation in $\Gamma$
and $R$.  In the absence of dissipation and gravity, equation
(\ref{eq:energy}) is in fact another algebraic equation relating $\Gamma$
and $R$, thus, the two equations can be solved for $\Gamma(p_{\rm ext})$
using a numerical root finder.  Apart from reproducing the scaling
behaviors established in \S\ref{sec:properties}, this will enable us to
determine the terminal Lorentz factors and the length scales over which the
transitions between different dynamical phases occur.  Furthermore, we can
use the full dynamical model to investigate the evolution of such
observational quantities as polarization and synchrotron brightness.

In the absence of dissipation, the terminal Lorentz factor that can be
reached with such a jet is simply
\begin{equation}
  \Gamma_{\infty} \equiv \lim_{p_{\rm ext} \rightarrow 0} \Gamma =
  \Gamma_0
  \left(\frac{{\rho'}_0 c^2 + 2{{U_{\perp}}'}_0 + 4{{p'}_0}}{{\rho'}_0
      c^2}\right),
  \label{eq:terminalgammazerolambda}
\end{equation}
where subscripts 0 denote quantities evaluated at some arbitrary upstream
point.

This simple solution is no longer possible in the presence of dissipation,
which introduces a sink term into equation \,(\ref{eq:energy}).  As a
result, we have to use equation \,(\ref{eq:energylosses}) instead.  Once
the energy has been converted into particle pressure, it can be radiated
away as isotropic radiation, which will not affect the dynamics of the jet
any further (assuming that $p$ is dynamically unimportant).  If the energy
is stored as particle pressure, the pressure could eventually become
dynamically important (see \S\ref{sec:equipartition}).  Until that happens,
though, the two solutions are identical.  The terminal Lorentz factor is
always reduced (see \S\ref{sec:tradeoff}), but the acceleration efficiency
can be increased for $\zeta < -3/5$ (see \S\ref{sec:selfsimilar}).  We have
plotted the solution for the radiative case (the one case solvable
analytically) in Fig. \ref{fig:Gammaplot}.

{\plotfiddle{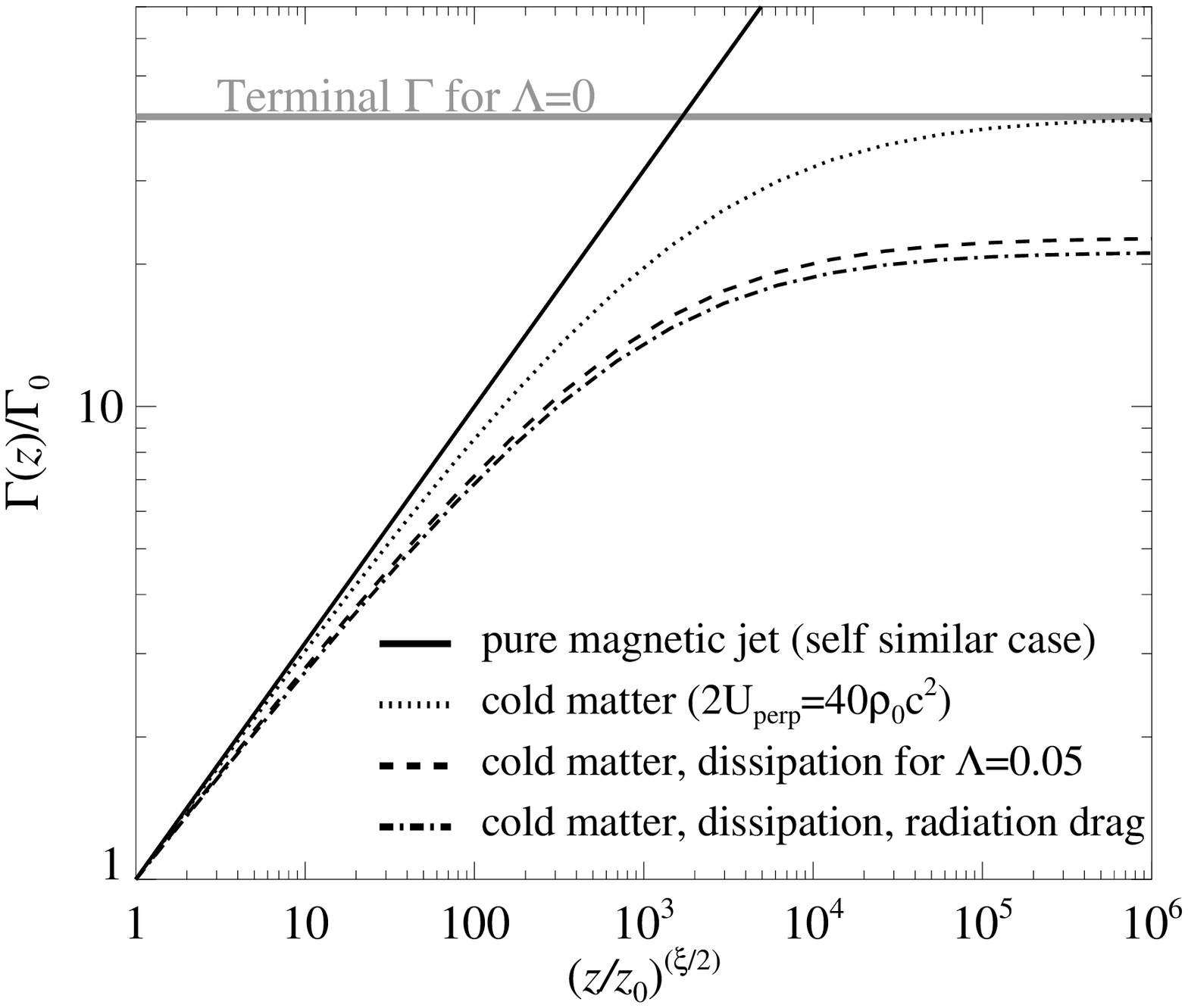}{3.0in}{0}{50}{50}{-128}{0} 
\figcaption{Analytic solutions in the limit $v \sim 1$ for $\zeta = 0$:
  a) self similar limit, valid for very large ${U_{\perp_0}} \gg \rho'
  c^2$; b) solution with cold matter, ${U_{\perp_0}}/\rho_0 c^2 = 20$,
  i.e., $\Gamma_{\infty} = 41$; c) radiative dissipation with $\Lambda
  = 0.05$, otherwise same parameters as b); d) same parameters as c)
  but including radiation drag.  Note that the dissipation of energy
  alters both the acceleration efficiency and the terminal Lorentz
  factor $\Gamma_{\infty}$.\label{fig:Gammaplot}}}
\bigskip


\subsection{Radiation Drag}
The presence of ultra-high-energy particles in AGN jets suggests that
inverse Compton (IC) enhanced radiation drag might be dynamically
important.  While O'Dell (1981) initially suggested that pair jets might be
accelerated by the Compton rocket effect, Phinney (1982) showed that it is
hard to accelerate a plasma beyond fairly modest Lorentz factors by
radiation pressure without a continuous source of particle acceleration to
offset the strong IC cooling of the plasma.  Furthermore, if the radiation
source is not point-like, the terminal Lorentz factor is limited by the
solid angle $\Omega$ the radiation source subtends.  On the other hand,
radiation drag can {\it hamper} the bulk acceleration of plasmas containing
relativistic particles in the presence of a radiation field, if those
particles are continuously reheated to overcome the IC losses.  The
dissipation mechanism discussed above could provide such reheating.  Here
we will consider the effect of radiation drag in the simplest possible
prescription.

We assume that the IC cooling and the dissipational heating time scales are
short compared to the adiabatic timescale.  If this is not satisfied, the
influence of radiation drag will be {\it reduced}.  We can then expect
dissipational heating to nearly balance IC losses in a near equilibrium
situation.  Thus the amount of IC drag is controlled by how much
dissipation there is.  For this approximation to be valid, IC losses must
dominate the loss processes of the particles, i.e., the radiation energy
density ${{U'}_{\rm rad}}$ in the comoving frame must be large compared to
the magnetic field energy density $U'$ (for large enough $\Gamma$ this is
always going to be the case, since the external field will be Doppler
boosted).  Finally, we assume that $\langle \beta^2\gamma^2\rangle \gg 1$,
where $\gamma$ is the particle Lorentz factor in the comoving frame.  This
sets an upper limit of 
\begin{eqnarray}
  {U'}_{\rm rad} & \ll & 6\times 10^{3}\,{\rm
    ergs\,cm^{-3}} \nonumber \\
  & & \ \ \cdot \left[\Gamma \Lambda \ \frac{U'}{\rho'
  c^2}\ \frac{d\ln{\Gamma 
  R^2}}{d\ln{z}}\ \frac{10^{14}{\, \rm
  cm}}{z}\right]
\end{eqnarray} 
on the comoving radiation energy density (otherwise IC cooling would have
lowered the upper spectral cutoff to $\gamma_2 \sim 1$).  These assumptions
allow us to eliminate ${U'}_{\rm rad}$ from the equations, since the drag
term and the cooling term are both proportional to ${U'}_{\rm rad}$.  In a
sense, then, we are presenting an upper limit on the importance of IC
radiation drag over large length scales.  It has to be kept in mind,
though, that drag can be much more important in non-stationary situations
(like, for example, in shocks), which are beyond the scope of this paper.

We assume the jet is moving through a radiation field that is locally
isotropic in the lab frame.  Following Phinney's treatment (1982), we can
calculate the loss rate and the force density due to IC scattering in the
comoving frame and then transform back to lab frame to find the additional
term for equation (\ref{eq:energy}).  We find that radiation drag always
decreases {\it both} the acceleration efficiency $\eta$ and the terminal
Lorentz factor $\Gamma_{\infty}$ by moderate amounts.  It does not,
however, introduce qualitatively new features.  To demonstrate this, we
have plotted a solution including radiation drag for otherwise identical
parameters in Fig.  \ref{fig:Gammaplot}.


\section{Discussion}
\label{sec:discussion}


\subsection{Tradeoff Between Dissipation and Acceleration and Synchrotron
  Brightness}
\label{sec:tradeoff}
In the following section we will investigate the observational effects of
the jets we have introduced in this paper.  A highly dissipative jet will
radiate away a large fraction of its internal energy along the way before
reaching the terminal Lorentz factor $\Gamma_{\infty}$, while a
non-dissipative jet will convert all its internal energy into kinetic
energy flux.  Since the jet will ultimately terminate and reconvert its
kinetic energy flux into random particle energy when it slams into the
surrounding medium, the ratio of kinetic luminosity (which could be
estimated based on the energy input into the lobes, based on the source
size and its age) to the radiative luminosity $L_{\rm diss}(z)$ (i.e., the
integrated luminosity of the jet before reaching the terminal shock) should
give us some indication of the importance of dissipation.

We have already seen in \S~\ref{sec:solutions} that the presence of
dissipation can lower $\Gamma_{\infty}$, thus lowering the kinetic energy
flux at the terminal shock, $L_{\infty}$ (dominated by cold kinetic energy
flux), and the produced hot-spot luminosity.  A given fraction $b$ of the
terminal luminosity $bL_{\infty}$ will be radiated away, which can be
estimated from the hot-spot and cocoon luminosity (calculating $b$ is, of
course, a highly non-trivial matter), giving us a handle on $\Lambda$.  We
have plotted the ratio
\begin{equation}
  \epsilon \equiv \frac{L_{\rm diss}}{L_{\infty}}
\end{equation}
in Fig. \ref{fig:TerminalGamma}.  To make that plot, we chose parameters
such that in the absence of dissipation the jet would reach
$\Gamma_{\infty}(\Lambda = 0) = {2{U'}_{\perp_0}}/({\rho'}_0 c^2) = 100$.
If we had chosen a larger (smaller) value of this parameter, the lines in
the plot would move down (up), since $\Gamma_{\infty}$ depends non-linearly
on both $\Lambda$ and ${2{U'}}_{0}/({\rho'}_0 c^2)$, so this plot is only a
representative one of a family of similar plots for different
$\Gamma_{\infty}(\Lambda = 0)$.

{\plotfiddle{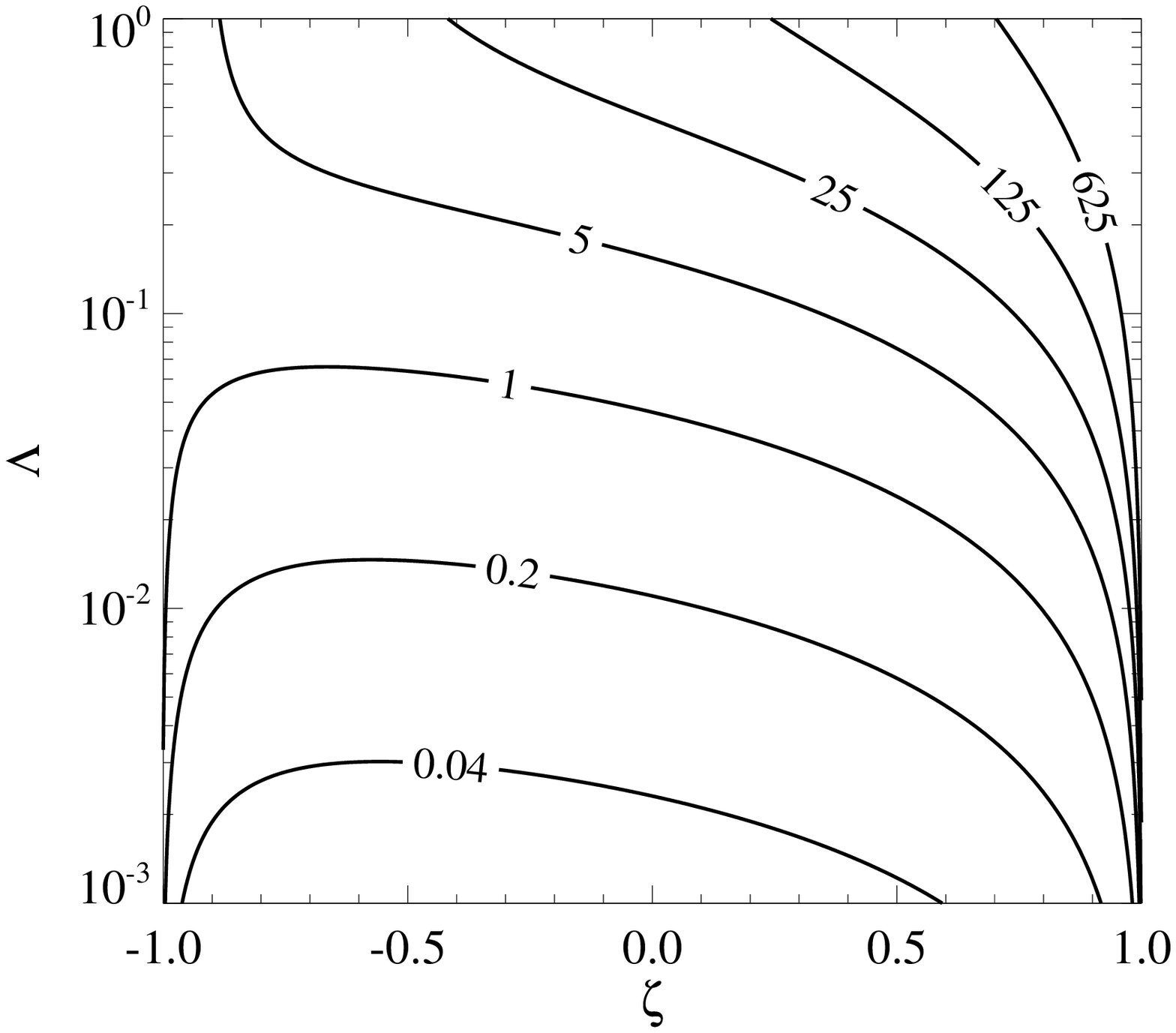}{3.0in}{0}{50}{50}{-128}{0}
\figcaption{Plot of the ratio $L_{\rm diss}/L_{\infty}$, indicating how
  much of the energy carried in the jet is radiated away and how much will
  reach the terminal shock (this energy can be used to heat the hot spot
  and to inflate the radio lobes).  This plot was constructed for
  ${\Gamma_{\infty_0}} \equiv \Gamma_{\infty}(\Lambda = 0) = 100$.  Lower
  values of ${\Gamma_{\infty_0}}$ will shift the lines in this plot
  upwards. Note that this plot was constructed using the assumption that
  all the dissipated energy is radiated away on the
  spot.\label{fig:TerminalGamma}}}
\bigskip

Another question is what the spatial distribution of the jet emission is,
since there are several competing effects: the optical depth to self
absorption (and inverse Compton up-scattering), Doppler boosting and
opening angle (the larger of which will determine the opening angle of the
cone into which most of the radiation goes), and of course the dissipative
power of the jet itself [depending on $\Lambda$ and $\Gamma(z)$, $R(z)$].
This question can be asked with respect to the {\it frequency integrated
  brightness} $I$ or the {\it spectral brightness} $I_{\nu}$.  If we assume
that all the dissipated energy is radiated away on the spot, we can
calculate the local dissipation rate, which must then be equal to the local
frequency integrated jet emissivity $j'$ in the comoving frame.  This
assumption depends on the injected particle energy spectrum.  If the
spectrum is flatter than $s = 2$, most of the energy is in the high energy
particles.  In that case, synchrotron cooling can be efficient enough for
our assumption of on the spot radiation to be effective.  If, on the other
hand, $s > 2$, most of the energy is in the low energy particles,
synchrotron radiation will not be efficient (unless $\gamma_1$ is very
high, in which case the injected spectrum would rapidly cool to a quasi
mono-energetic distribution), and the jet will accumulate particle pressure
or radiate by other means (note that Compton cooling would be equally
insufficient to balance heating in this case).

To investigate the first case we will set $s < 2$.  Given a viewing angle
we can then determine the observed total intensity $I$, given by
\begin{equation}
  I \propto \left.\left(\frac{d{U'}}{dz}\right)\right|_{\rm diss} D^2
  \frac{R}{\sin{\theta}},
\end{equation}
where $\theta$ is the angle between line of sight and jet axis and $D
\equiv \left[\Gamma\left(1 - v \cos{\theta}\right)\right]^{-1}$ is the
Doppler factor.  This expression takes relativistic beaming and the
relativistic corrections to foreshortening into account.  One might expect
that the integrated brightness peaks at a certain distance from the core,
since $D$ is strongly peaked at $\Gamma \sim 1/\theta$.  However, in our
prescription the dissipation drops off too fast for this effect to be
important.  The main difference in the brightness evolution is that a
dissipative jet has a different efficiency $\eta$ and generally expands
less rapidly in the sideways direction (and will reach a smaller terminal
Lorentz factor $\Gamma_{\infty}$).  This effect will become important once
the jet has reached $\Gamma_{\infty}$ and only for $\Lambda$ large enough
to significantly alter the dynamics ($\Lambda \gtrsim 0.1$).  We have
plotted $I$ as a function of $z$ for different values of $\Lambda$, $\zeta
=0$, ${U'}_0 = 20\,{\rho'}_0 c^2$, and a viewing angle of $\theta =
10^{\circ}$ in Fig.~\ref{fig:Intensity}, arbitrarily normalized to
$I(\Lambda = 0.01)$ to increase dynamic range (the brightness decreases by
many orders of magnitude along the jet).  As is obvious from the plot, for
small $\Lambda$, only the overall normalization of $I$ varies with
$\Lambda$, whereas for large enough $\Lambda$, the brightness distribution
itself changes shape due to the altered dynamics.  Also shown are the
Doppler factors $D$ for the different parameters, which are primarily
responsible for the different shapes.

The situation can become more difficult in the opposite case, i.e., if most
of the radiation is trapped (e.g., by synchrotron self-absorption, which
implies a steep spectrum, $s > 3$) or if the deposited energy is simply not
efficiently radiated (e.g., if $2 < s < 3$).  In that case the emission of
dissipated energy could be delayed, leading to a relative brightness peak
downstream.  To briefly investigate this possibility, we assume the latter
case, i.e., $2 < s < 3$, which is not an unreasonable choice for AGNs (see
\S\ref{sec:applications}, for example).  We assume that the energy flux
$\nu F_{\nu}$ peaks at high energies $\nu_{\rm p}$: either at the spectral
break of $\Delta \alpha \sim 1/2$ expected in a scenario in which high
energy particles are constantly re-injected, where the spectral index
$\alpha$ is given by
\begin{equation}
  \alpha \equiv -\frac{d\ln{I_{\nu}}}{d\ln{\nu}},
\end{equation}
or at the spectral cutoff produced by synchrotron and IC cooling (in the
absence of a strong break).  The position of $\nu_{\rm p}$ depends on
adiabatic effects, radiative cooling, and heating due to dissipation.  The
spectrum will be self-absorbed at low frequencies, which generally leads to
an observed spectral index of $\alpha \sim -5/2$ (for an exact treatment of
the spectral shape at the self absorption turnover see De Kool, Begelman,
\& Sikora 1989).  If we take $2 < s < 3$ or $1/2 < \alpha < 1$, the self
absorbed part contributes a negligible fraction to the total brightness and
most of the energy is emitted at the high end of the spectrum.

In this case it is impossible to calculate the brightness analytically as a
function only of $\Gamma$, $R$, and $z$.  Rather, one can numerically
integrate the evolution equation of the peak frequency under adiabatic
cooling, dissipative heating [we assume a self-similar transfer of energy
from magnetic field to the particles such that
$\left.\left(d\gamma/dz\right)\right|_{\rm diss} \propto \gamma
\left.\left(d{U'}/dz\right)\right|_{\rm diss}$], and synchrotron cooling on
the basis of the solution $\Gamma(z)$ given above.  We can estimate the run
of $I$ by scaling it with the brightness at $\nu_{\rm p}$, taking account
of relativistic beaming and aberration.  If we choose to normalize the
intensity curves as we did in the previous case, we can get around fixing
the absolute normalization of ${U'}$, since it only enters linearly into
the intensity and will thus cancel out upon normalization.  We used
${U'}_{\perp_0} = 20\,{\rho'}_0 c^2 $, along with $s = 5/2$ and plotted the
frequency integrated intensity in Fig.~\ref{fig:NonRadInten} with otherwise
the same parameter values as in Fig.~\ref{fig:Intensity}, once again
normalized to $I$ for $\Lambda = 0.01$.  Note that for large values of
$\Lambda$ our assumption that the jet be magnetically dominated and that
particle pressure be negligible can break down (see
Fig.~\ref{fig:GammaEq}), so curves with high $\Lambda$ are to be taken with
a grain of salt.  Nevertheless, it is interesting to note that the
intensity drops less rapidly with $z$ for larger values of $\Lambda$,
corresponding to the delayed emission mentioned above.  The shapes of these
curves depend only weakly on $s$.  The different bends in the curves stem
from the evolution of the Doppler factor (shown in
Fig.~\ref{fig:Intensity}), and the evolution of the peak frequency. The
slopes of the curves are produced mostly by the evolution of the lower
cutoff frequency $\gamma_1$ (see eq.~[\ref{eq:spectrum}]), which enters the
expression for the synchrotron brightness through the powerlaw
normalization of the particle distribution, and by the evolution of Lorentz
factor and jet radius (entering through the particle density and the
integration of the emissivity across the jet).

{\plotfiddle{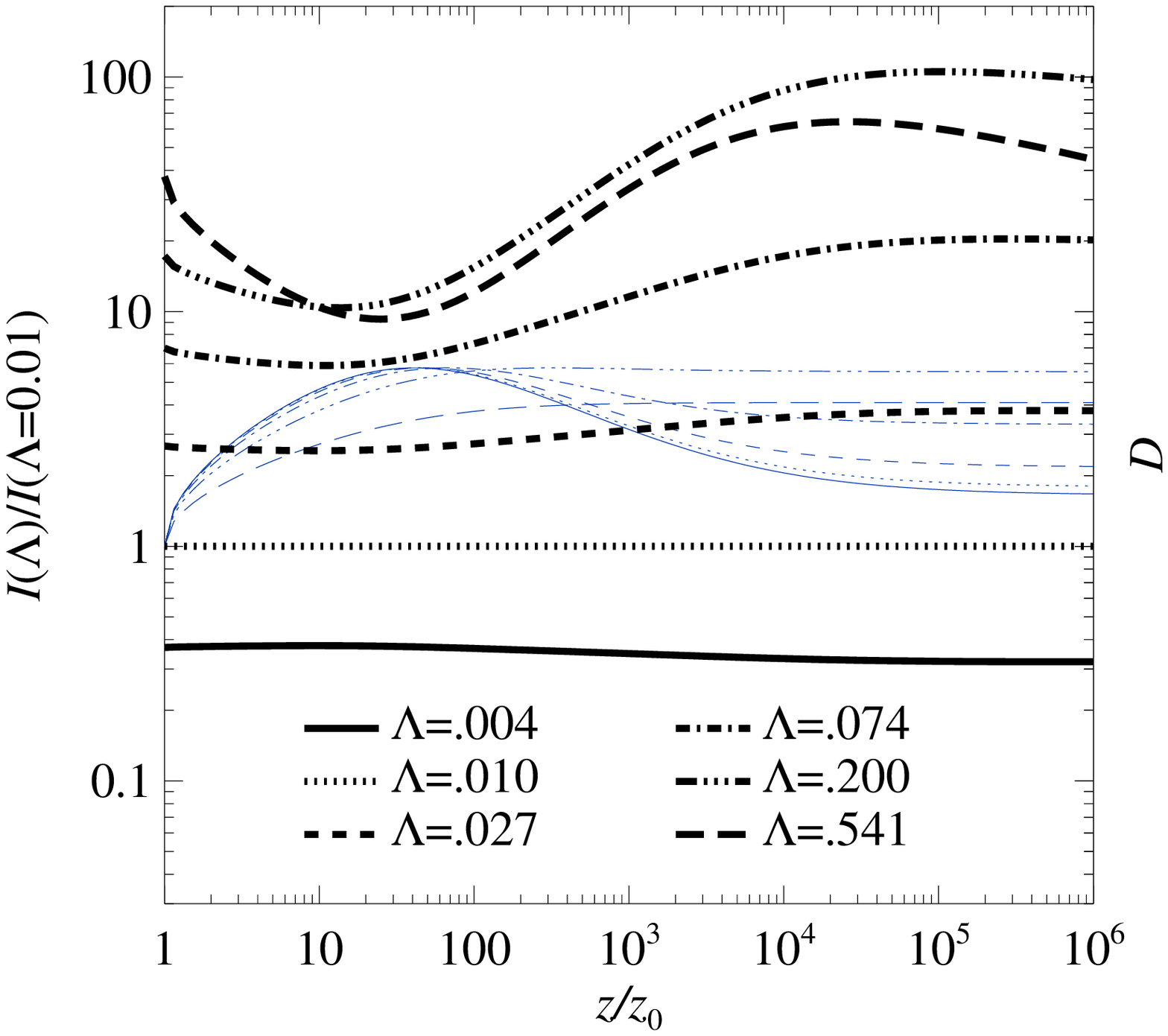}{3.0in}{0}{50}{50}{-128}{0} 
\figcaption{Plot of the frequency integrated brightness for a radiative jet
  (i.e., all the dissipated energy is radiated away on the spot) as a
  function of $z$ for different values of $\Lambda$, $\zeta = 0$,
  $\xi=2$, ${{U'}_{\perp_0}}/{\rho'_{0}} = 20$, and $\theta =
  10^{\circ}$, arbitrarily normalized to the intensity curve for
  $\Lambda = 0.01$ to increase the contrast (thick black curves).
  Also shown are the corresponding Doppler factors $D$ as thin grey
  curves.  The shapes of the individual intensity curves are mainly
  determined by the variation of $D$.  Note that a significant
  observable effect is achievable only for $\Lambda \gtrsim
  0.1$. \label{fig:Intensity}}}
\bigskip

{\plotfiddle{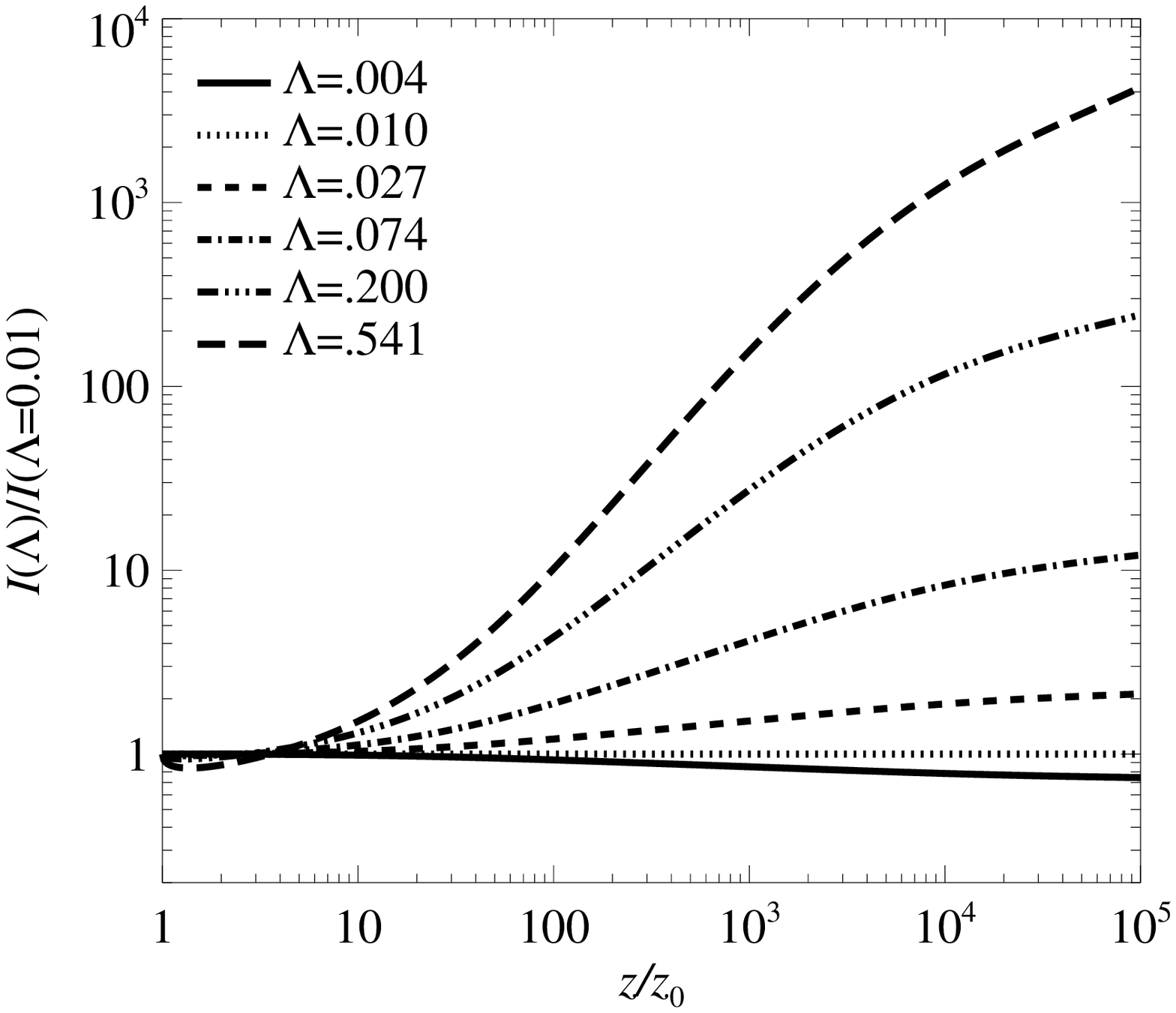}{3.0in}{0}{50}{50}{-128}{0}
\figcaption{Plot of the frequency integrated brightness for a marginally
  non-radiative jet (only a dynamically small amount of the dissipated
  energy is radiated away on the spot) as a function of $z$ for
  different values of $\Lambda$, $\zeta = 0$, ${U'}_0/{\rho'_{0} c^2}
  = 20$, $\xi=2$, and $\theta = 10^{\circ}$, arbitrarily normalized to
  the intensity curve for $\Lambda = 0.01$ to increase the contrast
  (the absolute intensity drops by many orders of magnitude). Note
  that these curves do not depend on the absolute value of the
  magnetic field, which cancels out due to the normalization.
  \label{fig:NonRadInten}}}
\bigskip

\subsection{Polarization}
\label{sec:polarization}
While the degree of polarization is highest for homogeneous magnetic
fields, jets with tangled or disorganized field can exhibit a net
polarization if there is a net anisotropy in the field (Laing 1980).  The
measured polarization will depend not only on $\zeta$, but also on the
viewing angle $\theta$ and the bulk Lorentz factor $\Gamma$.  The
polarization of radiation from a powerlaw distribution of electrons with
index $s$ in a region of homogeneous field is given by
\begin{equation}
  \Pi \equiv \frac{I_{\perp} - I_{\parallel}}{I_{\perp} + I_{\parallel}} =
  \frac{s + 1}{s + 7/3}
\end{equation}
where $I$ is the intensity at a given optically thin wavelength.  To
calculate the integrated polarization, we average across the jet. To do
this we decompose the radiation into polarization along the jet axis and
perpendicular to it.  Furthermore, we assume that field directions are
distributed among all solid angles and introduce a weighting function that
distributes the field orientations to the required anisotropy,
\begin{equation}
  w(\vartheta)={\sin{\vartheta}}^{\kappa}
  \label{eq:weighing}
\end{equation}
where $\vartheta$ is the angle between the field and the $z$-axis and
$\kappa$ is determined by the anisotropy.  We can solve for $\kappa$ under
the condition that $\langle {B_{\parallel}^2} \rangle = (1 + \zeta)/(1 -
\zeta)\langle{B_{\perp}}^2\rangle$:
\begin{equation}
  \kappa=-\frac{1 + 3\zeta}{1 + \zeta}.
\end{equation}
We correct the viewing angle for relativistic aberration, which has a
significant impact on the observed polarization, since the average
polarization will go to zero for a jet seen head on.  Furthermore, the
angle between line of sight and magnetic field is important in determining
the relative brightness of a region.  Figure \ref{fig:polarization} shows
the predicted polarization for the cases shown in Fig. \ref{fig:Gammaplot},
a spectral index of $\alpha = 0.5$, and a viewing angle of $\theta =
10^{\circ}$.  Since the anisotropy of the jet is fixed, the variation in
the polarization $\Pi(\theta)$ is solely caused by changes in the viewing
angle due to relativistic aberration.  Generally, the polarization will be
perpendicular to the jet axis if $\zeta < -1/3$ and parallel if $\zeta >
-1/3$.  As long as equation (\ref{eq:mixing}) holds, an extremum in
$\Pi(\theta)$ will be present and it should indicate the position where
$\Gamma = 1/\sin{\theta}$, i.e., where the viewing angle corrected for
aberration is $\theta' = 90^{\circ}$.  {Note that this polarization is
averaged across the jet.  In order to compare these predictions to actual
measurements a relatively small correction for the emission weighted
averaging across the jet at different angles must be made.  The qualitative
predictions of this section should be unaffected by that caveat.}

{\plotfiddle{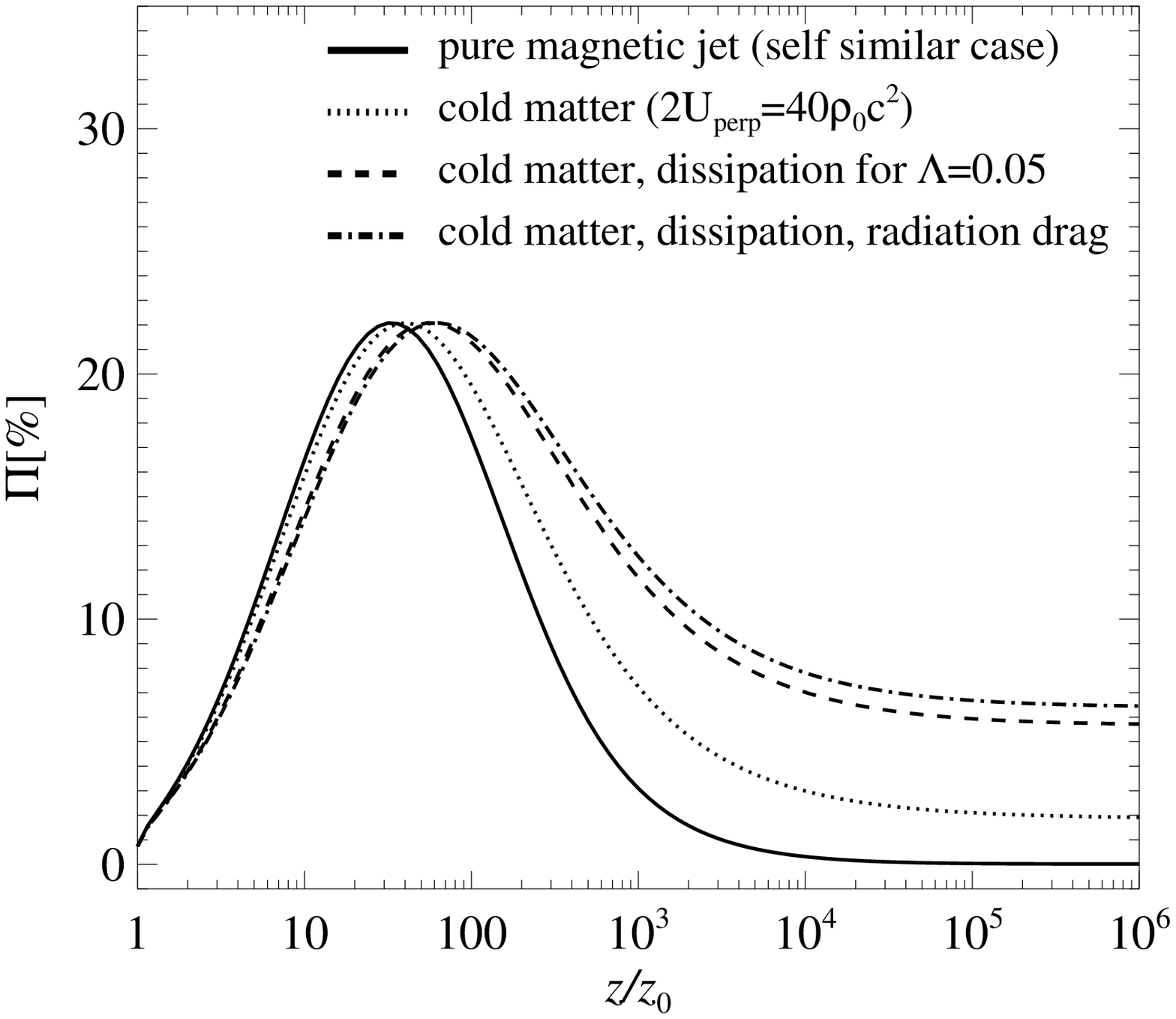}{3.0in}{0}{50}{50}{-128}{0}
\figcaption{Optically thin polarization $\Pi$ as a function of $z$ for the
  same parameter values as in Fig. \ref{fig:Gammaplot} and a viewing angle
  of $\theta = 10^{\circ}$.  We chose a spectral index of $\alpha
  = 0.5$.  The variation in $\Pi$ is solely due to changes in the aberrated
  viewing angle along the jet.\label{fig:polarization}}}
\bigskip


\subsection{Applications}
\label{sec:applications}
Finally, we sketch out some applications for this model.  The obvious
candidates for jet models are radio galaxies and all related jet powered
AGNs.  The best studied example is M87, since it is the closest unobscured
source (though by no means a particularly powerful one).  The central black
hole has a relatively well determined mass of $M \sim 2.4\times
10^{9}\,{\rm M_{\odot}}$ or $r_{\rm g} \sim 3.5\times 10^{14}\,{\rm cm}$.
The kinetic power of M87 is estimated to be of order $L_{\rm kin} \sim
10^{43 - 44} \,{\rm ergs\,sec^{-1}}$ (Reynolds {\it et al.}  1996, Bicknell
\& Begelman 1996).  Taking the new proper motion measurements based on HST
data (Biretta, Sparks, \& 1999) at face value, the terminal bulk Lorentz
factor likely falls into the range $6 \lesssim \Gamma_{\infty} < 10$ with a
viewing angle of $\theta \sim 20^{\circ}$. [Note that VLA measurement show
slower proper motions (Biretta, Zhou, \& Owen 1995) and the relationship
between the observed pattern speeds and $\Gamma$ is not known. There are,
however, no direct measurements of bulk motions, so for lack of better
knowledge we will use the larger values obtained from the optical data.]
The average polarization at kpc distances (where the jet has likely reached
terminal velocity) is roughly $\Pi \sim 10\%$ parallel to the jet axis ,
which corresponds to $\zeta \sim 0.3$ with the numbers given above.

Reynolds {\it et al.} (1996) showed that the jet probably consists of pairs
rather than ionized gas, so the jet might actually still be accelerating
(if indeed a large fraction of the particles has $\gamma > 10$. Note,
however, that the presence of shocks, clearly visible as knots in all
wavelengths, calls for a more sophisticated model which takes
time-dependence and MHD instability effects into account.  For a discussion
of the nature of the shocks observed in M87 see Bicknell \& Begelman 1996.)
Furthermore, there is evidence that the jet is {\it not} magnetically
dominated on large scales ($z > 100\,{\rm pc}$) and that the magnetic field
is actually somewhat below equipartition (Heinz \& Begelman 1997), which
means in this context that the magnetic field must have been dissipated
non-radiatively, accounting for the particle pressure at larger distances.
The observed spectral index at optically thin wavelengths is $\alpha \sim
0.6$, which is at least in the correct regime for synchrotron radiation to
be inefficient at radiating away the dissipated energy (see
\S\ref{sec:tradeoff}).

At a viewing angle of $20^{\circ}$ the jet radius at knot A ($z \sim
3\,{\rm kpc}$) is $R\sim 35 \,{\rm pc}$ with an approximate opening angle
of $\alpha_{\rm o} \sim 0^{\circ}.7$, much smaller than the beaming angle
of $\alpha_{\rm b} \gtrsim 6^{\circ}$, thus the jet is narrow at least at
VLA resolution.  Note that entrainment of ambient material may be
important before the jet reaches knot A (Bicknell \& Begelman 1996).
Using knot D as a reference point does not change this analysis
significantly, however.  There are no good estimates of the pressure
gradient in the innermost regions of the M87 X-ray atmosphere (this
situation should change, though, with the launch of {\it Chandra}).  The
pressure at large distances is approximately $p_{\rm ISM} \sim
10^{-10}\,{\rm dyn\,cm^{-2}}$, but Bicknell and Begelman (1996) argue that
the pressure in the radio lobes is significantly higher than this, $p_{\rm
  ext} \sim 1.5\times 10^{-9} \,{\rm dyn\, cm^{-2}}$.  For lack of better
information we will assume the latter value and a smooth pressure gradient
with $\xi = const$.  The radiative luminosity of the jet itself is $L_{\rm
  rad} \gtrsim 3 \times 10^{42}\,{\rm ergs\,sec^{-1}}$ (Reynolds {\it et
  al.}  1996), which argues for $\Lambda > 4\times 10^{-3}$.  If (as we
speculated above) much of the particle pressure at large distances was
indeed produced by dissipation, $\Lambda$ could be much larger.

The jet probably reaches terminal $\Gamma_{\infty} \sim 10$ somewhere
between VLBI scales and VLA scales, $2\times 10^{3}\, r_{\rm g} \sim
0.2\,{\rm pc} < z_{\infty} < 200\,{\rm pc} \sim 2 \times 10^{6}\, r_{\rm
  g}$.  For $\Lambda$ small enough to be dynamically unimportant we can
assume that $\eta \sim 1/4$ in the self similar regime.  Arbitrarily
setting $z_{0} \sim 10\, r_{\rm g}$ gives $0.8 \lesssim \xi \lesssim 1.74$.
The central pressure $p_0$ is then $2 \times 10^{-4}\,{\rm dyn\, cm^{-2}} <
p_{0} < 200\,{\rm dyn\, cm^{-2}}$, which requires an rms magnetic
field of $0.09\,{\rm G} < {B'}_{0} < 90\,{\rm G}$ for pressure balance
(using $\zeta \sim 0.3$).  The initial jet width strongly depends on $\xi$:
$8\, r_{\rm g} < R_0 < 9000\, r_{\rm g}$ for an assumed $z_0 = 10 r_{\rm
  g}$.  The latter value is unrealistic (and inconsistent with the limits
put on the jet width by VLBI observations, Junor \& Biretta 1995), since
most of the energy output of the disk into the jet will be provided close
to the black hole.  It is therefore most reasonable to assume that
$z_{\infty} \sim 2\times 10^{3}\, r_{\rm g}$ and $\xi \sim 1.7$.  The
total jet power implied by the numbers given is of order $L \sim 2 \times
10^{44}\,{\rm ergs\,sec^{-1}}$, consistent with the estimate by Bicknell \&
Begelman (1996).  Overall it seems that this model is consistent with the
observed properties of M87 to first order if we adopt a pressure gradient
following $\xi \sim 1.7$.

The total isotropic energy output of $E \gtrsim 10^{54}\,{\rm ergs}$ of GRB
990123 (Bloom {\it et al.} 1999, Bloom {\it et al.} 1999) argues strongly
in favor of non-isotropic gamma-ray burst (GRB) scenarios, so jet models
explaining the apparently beamed nature of these sources (e.g.,
M$\acute{\rm e}$sz$\acute{\rm a}$ros \& Rees 1997, Sari, Piran, \& Halpern
1999) enjoy newly enhanced popularity.  One of the standard scenarios for
the energy sources of GRBs is a massive accretion event (either a neutron
star - neutron star merger or the accretion of a neutron star by a black
hole: Narayan, Paczy$\acute{\rm n}$ski, \& Piran 1992), which leads to the
formation of a disk after tidal disruption of one of the objects.  Since
neutron stars already display large magnetic fields, one might expect
strong shear amplification of this field in the disk, leading to a scenario
similar to what we described in \S\ref{sec:model}.  Similarly, the
hypernova approach (Paczy$\acute{\rm n}$ski 1998, Woosley 1993) can produce
a precollimated outflow, which might then evolve into a jet.  Application
of our model to GRBs is, however, not as straightforward as in the case of
mature radio galaxies.  This is because GRBs are highly time dependent
(corresponding to the adolescent stages of radio galaxies): the lifetime of
the jet (roughly of the order of the light travel time of the material) is
shorter than the sound crossing time of the bubble the jet blows into the
environment, so a pressure balanced solution as described above (which will
be set up after the jet and the ambient material have equilibrated, i.e.,
after the jet has existed for a few sound crossing times) might not be a
good approximation.  The investigation of jet dynamics in the context of
GRBs and in the framework of tangled fields as introduced above will
therefore be the subject of another paper.  Here we simply wish to point
out the benefits of jet models in general and an approach based on our
model in particular:
\begin{itemize}
\item{Acceleration of GRB outflows by Poynting flux has the advantage that
    collimation can be provided not only by the external medium, but also
    by the field geometry itself (note that for $\delta = 0$ the
    perpendicular component of the field does not contribute to the
    sideways pressure, so for $\zeta \sim -1$ the jet can have a large
    Poynting flux yet orders of magnitude smaller sideways pressure than a
    particle dominated jet would have for the same energy flux).}
\item{One would not need to invoke shocks to produce emission in this
    context, since the internal dissipation of magnetic energy would
    provide a natural source of high energy particles and photons to
    produce gamma rays (see Thompson 1994 for an example of how internal
    dissipation can power GRBs).  The short term variability seen in GRBs
    could then be explained by inhomogeneities (e.g., variations in the
    inhomogeneity of the field) imprinted on the outflow by
    variability in the central engine itself, which is expected to have
    time scales of the same order as the ones observed.}
\end{itemize}


\section{Conclusions}
\label{sec:conclusions}
We have presented simple analytic models of jet acceleration.  The jets are
accelerated by tangled magnetic fields, with collimation being provided by
pressure from an external medium.  This is a new approach, based on
Begelman (1995), as previous models of MHD jet acceleration concentrated
mostly on large scale organized fields.  Our analytical quasi-1D approach
is limited to narrow jets, for which the opening angle is smaller than the
beaming angle, coincident with the condition that the jet be in causal
contact with its environment.  We introduce an ad-hoc process that
redistributes energy between perpendicular and parallel field to facilitate
efficient conversion of internal energy to kinetic energy.  Without such a
process, stationary jet acceleration by disorganized fields is impossible.
In the absence of dissipation, the rate of acceleration achieved under this
scenario is the same as in the case of relativistic particle pressure,
$\Gamma \propto {p_{\rm ext}}^{-1/4}$.  We also find analytic solutions
beyond the self-similar region, which enable us to calculate the terminal
bulk Lorentz factors of such jets.  In order for these jets to reach the
observed $\Gamma > 10$, they initially must be magnetically dominated.

We estimate the impact of dissipation of magnetic energy on the dynamics of
the flow by considering a simple, phenomenological prescription of the loss
process.  The presence of dissipation lowers the terminal Lorentz factor
$\Gamma_{\infty}$ and generally changes the rate at which the jet is
accelerated (the latter effect is noticeable only if the dissipation rate
is comparable to the adiabatic expansion rate).  We also include the
effects of radiation drag in the simplest scenario, which always
lowers the efficiency $\eta$ and $\Gamma_{\infty}$.  The amount of
radiation drag in our model is controlled by the amount of dissipation
replenishing the high energy particle pool but seems to be dynamically
unimportant.

We calculate the frequency integrated surface brightness for the extreme
cases where all or very little of the dissipated energy is radiated away on
the spot and find that, while the brightness drops off very rapidly in all
considered cases due to the expansion of the jet, values of the dissipation
efficiency $\Lambda \gtrsim 0.05$ can have a significant impact on the
intensity as a function of $z$.  In the marginally non-radiative case the
buildup of particle energy leads to a slower decline in intensity with $z$
for larger $\Lambda$.  Finally, we apply this model to the prototypical
radio galaxy M87 and find that it is consistent with the observed
properties.

\acknowledgements

This research was supported in part by NSF grants AST95--29170 and
AST98--76887. MCB also acknowledges support from the Institute for
Theoretical Physics under NSF grant PHY94--07194, and from a Guggenheim
Fellowship.  We thank Chris Reynolds, Jim Chiang, and Chris Thompson for
helpful discussions.

\begin{appendix}
  \section{A More Realistic Dissipation Law}
  The dissipation law we assumed in \S\ref{sec:dissipation} was only one of
  many plausible {\it ad hoc} models.  Since (for finite conductivity,
  i.e., beyond the limit of perfect MHD) reconnection will occur whenever
  there is field reversal on sufficiently small scales (which would
  certainly be the case in a highly tangled geometry), we would expect
  dissipation to occur even if there were no change in field geometry due
  to expansion or acceleration.  In that case we might expect that the
  dissipation timescale is proportional to the time it takes a disturbance
  to travel a given characteristic length (e.g., the jet width) in the
  comoving frame, i.e.,
  \begin{equation}
    \left.\frac{d{{U'}_{\rm i}}}{dz}\right|_{\rm diss} \sim
        \Lambda{U'}_{\rm i}\frac{v_{\rm
        Alfv\acute{e}n}}{\Gamma R},
    \label{eq:altdiss}
  \end{equation}
  where the parameter $\Lambda$ absorbs the effects of resistivity and all
  the unknown physics of the reconnection process. Once again, it is
  straightforward to generalize to the case of different $\Lambda_{\rm i}$
  for different components of the field.
  
  It is possible to solve the set of equations in the magnetically
  dominated case (i.e., setting $\rho' + 4p' =0$), under the assumption
  that $\zeta = const.$ and in the relativistic limit, $\Gamma \gg 1$.  In
  this limit $v_{\rm Alfv\acute{e}n} = 1$, which simplifies the treatment
  significantly.  We assume that the dissipated energy is radiated away
  immediately, which leads to a modified equation (\ref{eq:energylosses})
  \begin{equation}
    \frac{\Lambda}{\Gamma R}\left(\frac{2 + 6\zeta}{3 - 3\zeta}\right) +
    \left(2 + 2\zeta\right)\frac{d\Gamma}{\Gamma dz} - \left(2 +
      2\zeta\right)\frac{dR}{Rdz} = 0.
  \end{equation}
  The pressure balance equation (\ref{eq:pressure}) is also modified:
  \begin{equation}
    \left(\zeta - 1\right) \frac{d\Gamma}{\Gamma dz} - \left(3 +
      \zeta\right)\frac{dR}{Rdz} - \frac{\Lambda}{\Gamma R} = -\xi/z
  \end{equation}
  where we used ${U'} \propto p_{\rm ext}$ and equation \,(\ref{eq:xi}).
  This set of equations can be solved to give
  \begin{equation}
    \Gamma(z) = \Gamma_0 \left(z/z_0\right)^{\xi/4}\left\{1 - A
      \left[\left(z/z_0\right)^{1 - \xi/2} -
        1\right]\right\}^{\left(3 +5\zeta\right)/\left(4 + 4\zeta\right)},
  \end{equation}
  with
  \begin{equation}
    A \equiv \frac{\Lambda}{\Gamma_0 R_0}\frac{1}{(1 - \xi/2)(1 + 3\zeta)}.
  \end{equation}
  Subscripts $0$ denote quantities evaluated at some arbitrary upstream
  point $z_0$.  In the limit of $\Lambda \rightarrow 0$ this solution
  appropriately reduces to the result without dissipation, i.e., $\Gamma
  \propto z^{\xi/4}$.  The solution can essentially take on three different
  behaviors: If $\xi > 2$, the solution will asymptotically approach
  $\Gamma \propto z^{\xi/4}$, i.e., $\eta = 1/4$.  If $\xi < 2$, on the
  other hand, two different scenarios can occur: the flow can either
  approach a self-similar behavior with $\eta \not= 1/4$ (but still
  constant) in the limit of $z \gg z_0$, or, if $A > 0$ (i.e., $\zeta >
  -1/3$), the flow can actually stall, i.e., $\Gamma \rightarrow 0$ for $z
  \rightarrow z_0\left[\left(1 - \xi/2\right)\left(1 + 3\zeta\right)/A -
    1\right]^{1/(1 - \xi/2)}$.  In this limit, the magnetic field
  dissipates away too quickly to satisfy pressure balance and the jet must
  contract and decelerate to increase its internal pressure, thereby
  increasing its dissipation rate, which leads to a run-away process.
  Eventually, the massless aproximation will break down, in which case the
  Alfv$\acute{\rm e}$n velocity will drop, lowering the dissipation rate
  (furthermore, the particle pressure will gain in importance, eventually
  stabilizing the jet against external pressure, in which case the jet
  would behave as described by BR74).  This process would produce an
  observable hot-spot (and possibly a shock) at a fixed distance.
\end{appendix}
\end{document}